\documentclass[aps,prd,twocolumn,showpacs]{revtex4-1}

\usepackage{wallpaper}
\usepackage[colorlinks,linkcolor=blue,anchorcolor=blue,citecolor=red]{hyperref}
\usepackage{amsmath}
\usepackage{multirow}

\begin{document}
\hyphenpenalty=6000
\tolerance=1000

\hyphenation{Eq}

\title{Unified nuclear matter EOSs constrained by the in-medium balance in density-dependent covariant density functionals}

\author{Cheng-Jun~Xia$^{1,2,3}$}
\email{cjxia@yzu.edu.cn}
\author{Bao Yuan Sun$^{4, 5}$}
\email{sunby@lzu.edu.cn}
\author{Toshiki Maruyama$^{3}$}
\email{maruyama.toshiki@jaea.go.jp}
\author{Wen-Hui Long$^{4, 5}$}
\email{longwh@lzu.edu.cn}
\author{Ang Li$^{6}$}
\email{liang@xmu.edu.cn}

\affiliation{$^{1}${Center for Gravitation and Cosmology, College of Physical Science and Technology, Yangzhou University, Yangzhou 225009, China}
\\$^{2}${School of Information Science and Engineering, NingboTech University, Ningbo 315100, China}
\\$^{3}${Advanced Science Research Center, Japan Atomic Energy Agency, Shirakata 2-4, Tokai, Ibaraki 319-1195, Japan}
\\$^{4}${School of Nuclear Science and Technology, Lanzhou University, Lanzhou 730000, China}
\\$^{5}${Frontiers Science Center for Rare Isotopes, Lanzhou University, Lanzhou 730000, China}
\\$^{6}${Department of Astronomy, Xiamen University, Xiamen 361005, China}
}

\date{\today}

\begin{abstract}
Considering the effects of charge screening, we propose a new numerical recipe within the framework of Thomas-Fermi approximation, where the properties of nuclear matter throughout a vast density range can be obtained self-consistently. Assuming spherical and cylindrical approximations for the Wigner-Seitz cell, typical nuclear matter structures (droplet, rod, slab, tube, bubble, and uniform) are observed. We then investigate the EOSs and microscopic structures of nuclear matter with both fixed proton fractions and $\beta$-equilibration, where two covariant density functionals DD-LZ1 and DD-ME2 are adopted. Despite the smaller slope $L$ of symmetry energy obtained with the functional DD-LZ1, the curvature parameter $K_\mathrm{sym}$ is much larger than that of DD-ME2, which is attributed to the peculiar density-dependent behavior of meson-nucleon couplings guided by the restoration of pseudo-spin symmetry around the Fermi levels in finite nuclei. Consequently, different mass-radius relations of neutron stars are predicted by the two functionals. Different microscopic structures of nonuniform nuclear matter are obtained as well, which are expected to affect various physical processes in neutron star properties and evolutions.
\end{abstract}

\maketitle

\section{\label{sec:intro}Introduction}

The equation of state (EOS) for dense stellar matter was shown to play important roles in the properties of cold neutron stars, the evolution of proto-neutron stars, the dynamics of core-collapse supernovae, the formation of black holes, and the binary neutron star mergers~\cite{Pons1999_ApJ513-780, Horowitz2004_PRC69-045804, Lattimer2012_ARNPS62-485, Janka2012_ARNPS62-407, Bauswein2012_PRD86-063001, Rueda2014_PRC89-035804, Qi2016_RAA16-008, Watanabe2017_PRL119-062701, Sotani2019_MNRAS489-3022, Koeppel2019_ApJ872-L16, Baiotti2019_PPNP109-103714, Schuetrumpf2020_PRC101-055804, Bauswein2020_PRL125-141103, Gittins2020_PRD101-103025, Preau2021_MNRAS505-939}. Nevertheless, due to the lack of understanding for strongly interacting matter at large densities, there are still large ambiguities on the compositions and structures of dense stellar matter, which leads to the uncertainties in the corresponding EOSs~\cite{Xia2020_PRD102-023031, LI2020}. In particular, even for the cases without involving any new degrees of freedom (heavy baryons, mesons, or quarks), the uncertainties in the nuclear energy density functional is still significant at large densities and isospin asymmetries~\cite{Dutra2012_PRC85_035201, Dutra2014_PRC90-055203, Hebeler2021_PR890-1}, which in turn affects our understanding on the properties and dynamic evolutions of compact stars.

Since it is still challenging to simulate dense matter with lattice QCD, a current viable strategy is to constrain the properties of nuclear matter based on both nuclear and astrophysical studies. For example, according to various terrestrial experiments and nuclear theories, the nuclear matter properties around the saturation density ($n_0\approx 0.16\ \mathrm{fm}^{-3}$) are well constrained with the binding energy $B\approx -16$ MeV, the incompressibility $K = 240 \pm 20$ MeV~\cite{Shlomo2006_EPJA30-23}, the symmetry energy $S = 31.7 \pm 3.2$ MeV and its slope $L = 58.7 \pm 28.1$ MeV~\cite{Li2013_PLB727-276, Oertel2017_RMP89-015007}. A recent measurement with PREX-II suggests that the neutron skin thickness for $^{208}$Pb is $\Delta R_{np}=0.283\pm 0.071$ fm, which yields $L = 106 \pm 37$ MeV~\cite{PREX2021_PRL126-172502}. Meanwhile, the observation of two-solar-mass pulsars~\cite{Demorest2010_Nature467-1081, Antoniadis2013_Science340-1233232, Fonseca2016_ApJ832-167, Cromartie2020_NA4-72, Fonseca2021_ApJ915-L12}, the simultaneous measurements of the masses and radii for PSR J0030+0451 and PSR J0740+6620 via pulse-profile modeling~\cite{Riley2019_ApJ887-L21, Riley2021_ApJ918-L27, Miller2019_ApJ887-L24, Miller2021_ApJ918-L28} have put strong constraints on the EOSs of dense stellar matter.  The multi-messenger observations of the binary neutron star merger event {GRB} 170817A-{GW}170817-{AT} 2017gfo have placed the tidal deformability of $1.4 M_{\odot}$ neutron star within $70\leq \Lambda_{1.4}\leq 580$~\cite{LVC2018_PRL121-161101}. A combination of those constraints and the heavy ion collision data suggests $K = 250.23 \pm 20.16$ MeV, $S = 31.35 \pm 2.08$ MeV and $L = 59.57  \pm 10.06$ MeV~\cite{Zhang2020_PRC101-034303}, while including PREX-II and chiral effective field theory constraints yields $S = 33.0^{+2.0}_{-1.8}$ MeV and $L = 53^{+14}_{-15}$ MeV~\cite{Essick2021_PRL127-192701}.

Relativistic-mean-field (RMF) models~\cite{Meng2016_RDFNS} have been very successful in describing finite nuclei~\cite{Reinhard1989_RPP52-439, Ring1996_PPNP37_193-263,
Meng2006_PPNP57-470, Paar2007_RPP70-691, Meng2015_JPG42-093101, Meng2016_RDFNS, Chen2021_SCPMA64-282011, Typel1999_NPA656-331, Vretenar1998_PRC57-R1060, Lu2011_PRC84-014328} and nuclear matter~\cite{Glendenning2000, Ban2004_PRC69-045805, Weber2007_PPNP59-94, Long2012_PRC85-025806, Sun2012_PRC86-014305, Wang2014_PRC90-055801, Fedoseew2015_PRC91-034307, Gao2017_ApJ849-19}. In this work we thus adopt RMF model to investigate the properties of dense stellar matter. According to the self-energies obtained in Dirac-Brueckner calculations starting from realistic nucleon-nucleon interactions, the density-dependent nucleon-meson coupling constants were introduced in RMF model to account for the in-medium effects in nuclear matter~\cite{Typel1999_NPA656-331, Roca-Maza2011_PRC84-054309}, which bypasses the problems of stability at large densities typically observed in traditional nonlinear RMF models.

To improve the density dependent behavior and give a better extrapolation of nuclear matter properties at large densities and isospin asymmetries, a new RMF Lagrangian DD-LZ1 guided by the restoration of pseudo-spin symmetry (PSS) was recently developed~\cite{Wei2020_CPC44-074107}. The PSS corresponds to a quasi-degeneracy between the two single-particle orbitals ($n,l,j=l+1/2$) and ($n-1,l+2,j=l+3/2$) in finite nuclei~\cite{Hecht1969_NPA137-129, Arima1969_PLB30-517}, which is attributed to the relativistic symmetry with a delicate balance between the nuclear attractive (scalar) and repulsive (vector) potentials~\cite{Liang2015_PR570-1, Geng2019_PRC100-051301}. Although RMF models well accommodate the PSS observed in stable nuclei~\cite{Liang2015_PR570-1}, the PSS is often violated for the high-$l$ orbitals in the vicinity of the Fermi surface, leading to spurious shell closures at $N$/$Z=58$ and 92. This problem can be solved if the RMF Lagrangian DD-LZ1 was adopted, where new density-dependent meson-nucleon coupling strengths were introduced~\cite{Wei2020_CPC44-074107}. It was shown that both the bulk properties of nuclear matter and finite nuclei are well described by the new RMF Lagrangian DD-LZ1. Meanwhile, the predicted density slope of nuclear symmetry energy within DD-LZ1 agrees with the experimental constraints, but a little softer than DD-ME2. In this work, we thus adopt the covariant density functional DD-LZ1~\cite{Wei2020_CPC44-074107} to investigate the EOS of cold nuclear matter expecting a better extrapolation towards large densities and isospin asymmetries. To show the variations in adopting the new density-dependent meson-nucleon coupling strengths, the obtained results are then compared with those of DD-ME2~\cite{Lalazissis2005_PRC71-024312}.

Due to the lack of knowledge for the nuclear energy density functionals, the uncertainties in the EOSs and microscopic structures of neutron star matter are still significant~\cite{Douchin2001_AA380-151, Oyamatsu2007_PRC75-015801, Grill2012_PRC85-055808, Bao2015_PRC91-015807, Sharma2015_AA584-A103, Fortin2016_PRC94-035804, Pearson2018_MNRAS481-2994, Liu2018_PRC97-025801, Shen2020_ApJ891-148, Vinas2021_Symmetry13-1613, DinhThi2021_AA654-A114, Newton2021_arXiv:2112.12108}. Additional uncertainties will be introduced if the EOSs were not obtained in a unified manner~\cite{Fortin2016_PRC94-035804, DinhThi2021_AA654-A114}. Particularly, the importance of a consistent calculation of nuclear functional was emphasized in Ref.~\cite{DinhThi2021_AA654-A114}, where using a surface tension that is inconsistent with the bulk functional would lead to an underestimation of both the average values and the uncertainties in the pasta properties. In this work we thus investigate the EOSs of cold nuclear matter as well as the corresponding microscopic structures in a unified manner, where the surface, curvature, and bulk contributions are obtained self-consistently from one single covariant energy density functional. For the nonuniform structures of nuclear matter, we adopt Thomas-Fermi approximation (TFA) and search for the ground state among the five types of nuclear matter structures (droplet, rod, slab, tube, and bubble)~\cite{Maruyama2005_PRC72-015802, Avancini2008_PRC78-015802, Avancini2009_PRC79-035804, Okamoto2012_PLB713-284, Gupta2013_PRC87-028801, Okamoto2013_PRC88-025801, Xia2021_PRC103-055812}, where the spherical and cylindrical approximations were imposed. The properties of nuclear matter are obtained with RMF models adopting the covariant density functionals DD-LZ1~\cite{Wei2020_CPC44-074107} and DD-ME2~\cite{Lalazissis2005_PRC71-024312}.

The paper is organized as follows. In Sec.~\ref{sec:the_RMF}, we present our theoretical framework of RMF model. The numerical details on obtaining the EOSs and microscopic structures of nuclear matter are discussed in Sec.~\ref{sec:the_EOS}. The obtained results on the EOSs and microscopic structures of nuclear matter as well as the implications for the structures of neutron stars are presented in Sec.~\ref{sec:results}. Our conclusion is given in Sec.~\ref{sec:con}.

\section{\label{sec:the_RMF} RMF model}
In the mean field approximation (MFA), the Lagrangian density of RMF models~\cite{Meng2016_RDFNS} for systems with time-reversal symmetry reads
\begin{eqnarray}
\mathcal{L}
 &=& \sum_{i=n,p} \bar{\psi}_i
       \left[  i \gamma^\mu \partial_\mu - \gamma^0 \left(g_\omega\omega + g_\rho\rho\tau_i + A q_i\right)- M^* \right] \psi_i
\nonumber \\
 &&\mbox{} + \sum_{l=e,\mu} \bar{\psi}_l \left[ i \gamma^\mu \partial_\mu - m_l + e \gamma^0 A \right]\psi_l - \frac{1}{4} A_{\mu\nu}A^{\mu\nu}
\nonumber \\
 &&\mbox{} + \frac{1}{2}\partial_\mu \sigma \partial^\mu \sigma  - \frac{1}{2}m_\sigma^2 \sigma^2
           - \frac{1}{4} \omega_{\mu\nu}\omega^{\mu\nu} + \frac{1}{2}m_\omega^2 \omega^2
\nonumber \\
 &&\mbox{} - \frac{1}{4} \rho_{\mu\nu}\rho^{\mu\nu} + \frac{1}{2}m_\rho^2 \rho^2,
\label{eq:Lagrange}
\end{eqnarray}
where $\tau_i$ represents the 3rd component of isospin for nucleon $i$, $q_i$ the charge ($q_p=e$, $q_n=0$, $q_e=q_\mu=-e$), and $M^*\equiv M + g_{\sigma} \sigma$ the effective nucleon mass. The meson fields $\sigma$, $\omega$, and $\rho$ take mean values with the field tensors $\omega_{\mu\nu}$, $\rho_{\mu\nu}$, and $A_{\mu\nu}$ vanish except for
\begin{equation}
\omega_{i0} = -\omega_{0i} = \partial_i \omega,
 \rho_{i0}  = -\rho_{0i}   = \partial_i  \rho,
  A_{i0}    = -A_{0i}      = \partial_i A.\nonumber
\end{equation}
Based on the Typel-Wolter ansatz~\cite{Typel1999_NPA656-331}, the density dependence of the coupling constants $g_{\xi}~(\xi=\sigma$, $\omega$) and $g_{\rho}$ are obtained with
\begin{eqnarray}
g_{\xi}(n_\mathrm{b}) &=& g_{\xi}(n_0) a_{\xi} \frac{1+b_{\xi}(n_\mathrm{b}/n_0+d_{\xi})^2}
                          {1+c_{\xi}(n_\mathrm{b}/n_0+e_{\xi})^2}, \label{eq:ddcp_TW} \\
g_{\rho}(n_\mathrm{b}) &=& g_{\rho}(0) \exp{\left[-a_\rho(n_\mathrm{b}/n_0)\right]}, \label{eq:ddcp_rho}
\end{eqnarray}
where $n_\mathrm{b} = \sum_{i=n,p} n_i$ represents the baryon number density of nuclear matter with $n_0$ being the saturation density.

Carrying out standard variational procedure, the equations of motion are then determined by
\begin{eqnarray}
(-\nabla^2 + m_\sigma^2) \sigma + g_{\sigma} n_\mathrm{s} &=& 0, \label{eq:KG_sigma} \\
(-\nabla^2 + m_\omega^2) \omega - g_{\omega} n_\mathrm{b} &=& 0, \label{eq:KG_omega}\\
(-\nabla^2 + m_\rho^2) \rho - \sum_{i=n,p} g_{\rho}\tau_{i} n_i &=& 0, \label{eq:KG_rho}\\
 \nabla^2 A + e(n_p - n_e - n_\mu) &=& 0. \label{eq:KG_photon}
\end{eqnarray}
In this work we adopt TFA and consider only zero temperature cases, the local nucleon scalar and vector densities are then obtained with
\begin{eqnarray}
n_{s} &=& \sum_{i=n,p} \langle \bar{\psi}_i \psi_i \rangle = \sum_{i=n,p} \frac{{M^*}^3}{2\pi^2} g\left(\frac{\nu_i}{M^*}\right),\\
n_i &=& \langle \bar{\psi}_i\gamma^0 \psi_i \rangle = \frac{\nu_i^3}{3\pi^2},
\end{eqnarray}
where $\nu_i$ is the Fermi momentum and $g(x) = x \sqrt{x^2+1} - \mathrm{arcsh}(x)$. The total energy of the system is then fixed by
\begin{equation}
E=\int \langle {\cal{T}}_{00} \rangle \mbox{d}^3 r, \label{eq:energy}
\end{equation}
with the energy momentum tensor
\begin{eqnarray}
\langle {\cal{T}}_{00} \rangle
&=& \mathcal{E}_0
     + \frac{1}{2}(\nabla \sigma)^2 + \frac{1}{2}m_\sigma^2 \sigma^2
     + \frac{1}{2}(\nabla \omega)^2 + \frac{1}{2}m_\omega^2 \omega^2 \nonumber \\
&&   + \frac{1}{2}(\nabla \rho)^2 + \frac{1}{2}m_\rho^2 \rho^2
     + \frac{1}{2}(\nabla A)^2.
\label{eq:ener_dens}
\end{eqnarray}
Adopting no-sea approximation, the local kinetic energy density is determined by
\begin{equation}
\mathcal{E}_0 = \sum_{i} \frac {{m^*_i}^4}{8\pi^{2}} \left[x_i(2x_i^2+1)\sqrt{x_i^2+1}-\mathrm{arcsh}(x_i) \right],
\end{equation}
where $x_i\equiv \nu_i/m^*_i$ with $m_n^*=m_p^*\equiv M + g_{\sigma} \sigma$, $m_e^*=m_e = 0.511$ MeV, and $m_\mu^*=m_\mu = 105.66$ MeV.

By minimizing the total energy $E$ with respect to the density profiles $n_i$ at fixed total particle numbers $N_i=\int n_i \mbox{d}^3 r$, one obtains the ground state which follows the constancy of chemical potentials, i.e.,
\begin{equation}
\mu_i(\vec{r}) = \sqrt{{\nu_i}^2+{m_i^*}^2} + \Sigma^\mathrm{R} + g_{\omega} \omega + g_{\rho}\tau_{i} \rho + q_i  A = \rm{constant}, \label{eq:chem_cons}
\end{equation}
with the additional ``rearrangement" term
\begin{equation}
\Sigma^\mathrm{R}=
 \frac{\mbox{d} g_\sigma}{\mbox{d} n_\mathrm{b}} \sigma n_\mathrm{s}+
   \frac{\mbox{d} g_\omega}{\mbox{d} n_\mathrm{b}} \omega n_\mathrm{b}+
   \frac{\mbox{d} g_\rho}{\mbox{d} n_\mathrm{b}} \rho \sum_i\tau_i n_i
\label{eq:re_B}
\end{equation}
due to the density dependent coupling constants adopted here~\cite{Lenske1995_PLB345-355}.

\section{\label{sec:the_EOS} Numerical details}

Nuclear matter at various densities, temperatures, and isospin asymmetries exhibits at least two phases, i.e., the liquid and gas phases~\cite{Yang2019_PRC100-054314, Yang2021_PRC103-014304}. Cold neutron star matter with densities $n_\mathrm{b}\gtrsim 0.08\ \mathrm{fm}^{-3}$ is expected to be in a uniform liquid phase, which is typically found in the core region of traditional neutron stars. At $n_\mathrm{b}\lesssim 0.08\ \mathrm{fm}^{-3}$, coexistence of the liquid phase and neutron gas takes place and exhibits various nonuniform structures~\cite{Baym1971_ApJ170-299, Negele1973_NPA207-298, Ravenhall1983_PRL50-2066, Hashimoto1984_PTP71-320, Williams1985_NPA435-844}, which are usually referred to as nuclear pasta and comprise of the inner crust of a neutron star or the core of supernovae at the stage of gravitational collapse. Adopting spherical and cylindrical approximations for the Wigner-Seitz (WS) cell~\cite{Pethick1998_PLB427-7, Oyamatsu1993_NPA561-431, Maruyama2005_PRC72-015802, Togashi2017_NPA961-78, Shen2011_ApJ197-20}, aside from the uniform phase, five types of pasta structures were observed, i.e, droplet, rod, slab, tube, and bubble. Meanwhile, more complicated structures may emerge if the spherical and cylindrical approximations were not imposed~\cite{Oyamatsu1984_PTP72-373, Magierski2002_PRC65-045804, Watanabe2003_PRC68-035806, Newton2009_PRC79-055801, Nakazato2009_PRL103-132501, Okamoto2012_PLB713-284, Schneider2014_PRC90-055805,  Schuetrumpf2015_PRC91-025801, Sagert2016_PRC93-055801, Berry2016_PRC94-055801, Fattoyev2017_PRC95-055804, Schuetrumpf2019_PRC100-045806, Kashiwaba2020_PRC101-045804, Xia2021_PRC103-055812}. At densities smaller than neutron drip density ($n_\mathrm{b}\lesssim 0.0003\ \mathrm{fm}^{-3}$), the neutron gas vanish and neutron star matter are comprised of finite nuclei in Coulomb lattices, which form the outer crusts of neutron stars as well as white dwarfs.

We thus divide the current section into two parts, i.e., the uniform nuclear matter in Sec.~\ref{sec:EOS_uniform} and the nonuniform one with two different density regions in Sec.~\ref{sec:EOS_Nonuniform}. For both cases, the covariant density functionals DD-LZ1~\cite{Wei2020_CPC44-074107} and DD-ME2~\cite{Lalazissis2005_PRC71-024312} are adopted.  Note that for the nonuniform nuclear matter, the effects of charge screening was shown to affect the microscopic structures (shape, nuclear radius $R_d$, cell size $R_\mathrm{W}$, etc.) of nuclear pasta~\cite{Maruyama2005_PRC72-015802}, which are addressed in this work with the electrons move freely and fulfill the constancy of chemical potential in Eq.~(\ref{eq:chem_cons}).

\subsection{\label{sec:EOS_uniform} Uniform nuclear matter}
For uniform nuclear matter, the mean fields and densities are independent of the space coordinates, then the derivative terms in the Klein-Gordon equations~(\ref{eq:KG_sigma}-\ref{eq:KG_photon}) vanish. At given baryon number densities $n_\mathrm{b}$ and proton fractions $Y_p\equiv n_p/n_\mathrm{b}$, the meson fields are obtained by solving Eqs.~(\ref{eq:KG_sigma}-\ref{eq:KG_rho}) with the density dependent meson-nucleon couplings fixed by Eqs.~(\ref{eq:ddcp_TW}) and (\ref{eq:ddcp_rho}). The energy density and chemical potentials are then determined by Eqs.~(\ref{eq:ener_dens}) and (\ref{eq:chem_cons}). Note that the Coulomb interaction is neglected for infinite nuclear matter, i.e., we have assumed $e=0$ in Eq.~(\ref{eq:KG_photon}). For realistic neutron star matter, the Coulomb interaction ensures the fulfillment of local charge neutrality condition with the inclusion of leptons, i.e.,
\begin{equation}
  n_p - n_e - n_\mu\equiv 0.
\end{equation}

\begin{figure}[!ht]
  \centering
  \includegraphics[width=\linewidth]{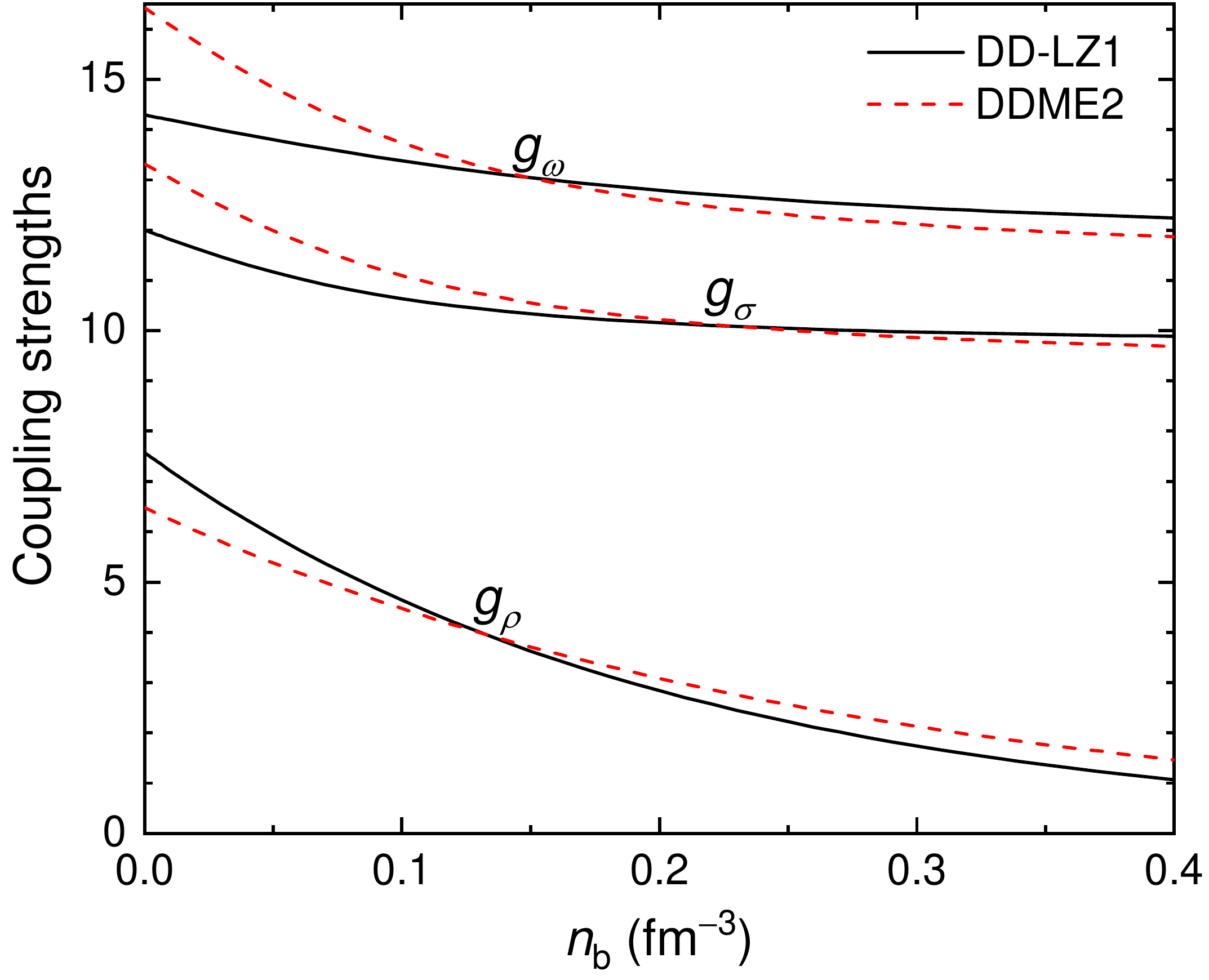}
  \caption{\label{Fig:Coupling} Meson-nucleon couplings as functions of baryon number density, which are adopted in the two covariant density functionals DD-LZ1~\cite{Wei2020_CPC44-074107} and DD-ME2~\cite{Lalazissis2005_PRC71-024312}.}
\end{figure}

The density dependence of the coupling strengths adopted in the covariant density functionals DD-LZ1~\cite{Wei2020_CPC44-074107} and DD-ME2~\cite{Lalazissis2005_PRC71-024312} are illustrated in Fig.~\ref{Fig:Coupling}. We note that $g_\sigma$ and $g_\omega$ obtained with DD-ME2 are in parallel with each other, which is typical in previous density-dependent RMF Lagrangians. For DD-LZ1, on the contrary, the variation of $g_\sigma$ with respect to density is smaller than that of $g_\omega$. Such a peculiar density dependent behavior is attributed to the restoration of PSS for high-$l$ orbitals~\cite{Wei2020_CPC44-074107}. Due to the enhanced centrifugal repulsion, the high-$l$ orbitals are usually located in the surface regions of finite nuclei, where the densities are much smaller than the center region and thus requires a different density dependent behavior for $g_\sigma$ and $g_\omega$ in order to reach PSS~\cite{Wei2020_CPC44-074107}. Meanwhile, as density increases, both $g_\sigma$ and $g_\omega$ of DD-LZ1 decrease slower than DD-ME2, while $g_\rho$ decreases slightly faster for DD-LZ1.

\begin{figure}[!ht]
  \centering
  \includegraphics[width=\linewidth]{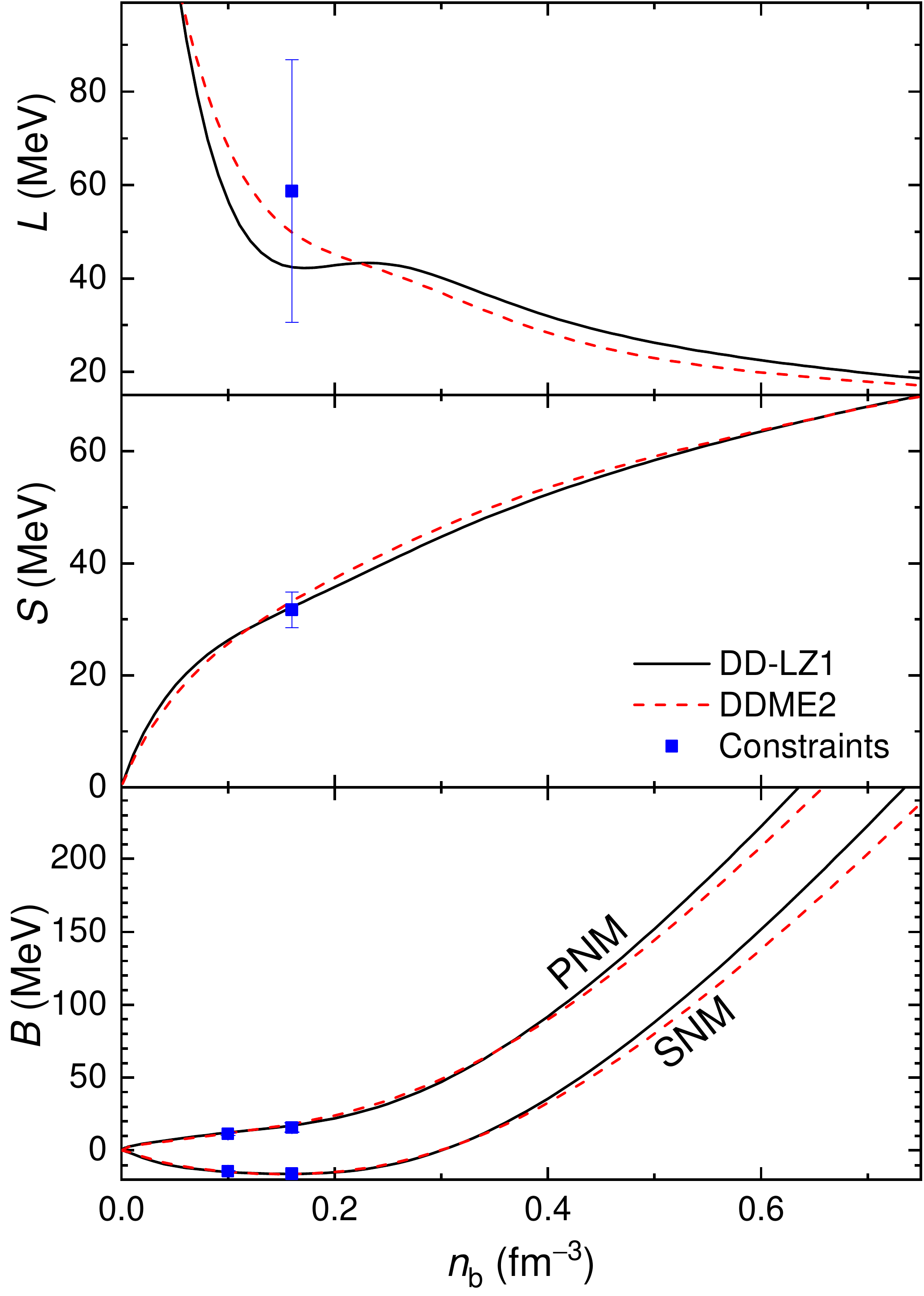}
  \caption{\label{Fig:Binding}Binding energy per nucleon for symmetric nuclear matter (SNM) and pure neutron matter (PNM), the corresponding symmetry energy $S(n_\mathrm{b})$ and its slope $L(n_\mathrm{b})$ as functions of baryon number density. The constrains  $B_\mathrm{PNM}(n_\mathrm{on}) = 11.4\pm1.0$ MeV, $B_\mathrm{SNM}(n_\mathrm{on}) = -14.1\pm0.1$ MeV~\cite{Brown2013_PRL111-232502}, $B_\mathrm{SNM}(n_0) = -16$ MeV, $B_\mathrm{PNM}(n_0) = B_\mathrm{SNM}(n_0) + S(n_0) = 15.7 \pm 3.2$ MeV, $S(n_0) = 31.7 \pm 3.2$ MeV, and $L(n_0) = 58.7 \pm 28.1$ MeV~\cite{Li2013_PLB727-276, Oertel2017_RMP89-015007} are denoted by the solid squares.}
\end{figure}

\begin{table}
\caption{\label{table:NM} The saturation properties of nuclear matter obtained with the covariant density functionals DD-LZ1~\cite{Wei2020_CPC44-074107} and DD-ME2~\cite{Lalazissis2005_PRC71-024312}. The corresponding root-mean-square deviations $\Delta$ from the experimental binding energies of finite nuclei~\cite{Audi2017_CPC41-030001, Huang2017_CPC41-030002, Wang2017_CPC41-030003} are presented as well, where the data are taken from Ref.~\cite{Wei2020_CPC44-074107}.}
\begin{tabular}{c|cccccc|c} \hline \hline
       & $n_0$        &   $B$    &   $K$  &  $S$    &  $L$  & $K_\mathrm{sym}$ &   $\Delta$       \\
       & fm${}^{-3}$  &   MeV    &   MeV  &   MeV   &  MeV  &   MeV            &    MeV   \\   \hline
DD-LZ1 &  0.158       & $-$16.06 &  230.7 &  32.0   &  42.5 &  $-20$           &   1.923  \\
DD-ME2 &  0.152       & $-$16.13 &  250.8 &  32.3   &  51.2 &  $-87$           &   2.400 \\
\hline
\end{tabular}
\end{table}

The differences in the density-dependence of meson-nucleon couplings lead to different predictions on the properties of nuclear matter as well as finite nuclei. In Table~\ref{table:NM} the saturation properties of nuclear matter corresponding to the covariant density functionals DD-LZ1~\cite{Wei2020_CPC44-074107} and DD-ME2~\cite{Lalazissis2005_PRC71-024312} are illustrated, while the root-mean-square deviations from the experimental binding energies of finite nuclei are presented as well~\cite{Wei2020_CPC44-074107}. It is evident that DD-LZ1 gives a better description for the binding energies of finite nuclei in comparison with DD-ME2. Meanwhile, the nuclear matter properties around the saturation density obtained by both functionals are consistent with the constraints $B\approx -16$ MeV, $K = 240 \pm 20$ MeV~\cite{Shlomo2006_EPJA30-23}, $S = 31.7 \pm 3.2$ MeV, and $L = 58.7 \pm 28.1$ MeV~\cite{Li2013_PLB727-276, Oertel2017_RMP89-015007}. The binding energy per nucleon for both pure neutron matter (PNM, $Y_p = 0$) and symmetric nuclear matter (SNM, $Y_p = 0.5$) are presented in Fig.~\ref{Fig:Binding}, while the corresponding symmetry energy and its slope are indicated as well. It is found that both functionals predict similar values for $B(n_\mathrm{b})$ and $S(n_\mathrm{b})$. Nevertheless, DD-LZ1 gives a peculiar density-dependent behavior for $L(n_\mathrm{b})$ with larger $K_\mathrm{sym}(n_0)$ in comparison with that of DD-ME2, which is mainly due to the novel density-dependent meson-nucleon coupling strengths adopted by DD-LZ1~\cite{Wei2020_CPC44-074107}. Note that at $n_\mathrm{on} = 0.1\ \rm{fm}^{-3}$ a robust constraint was found with $B_\mathrm{PNM}(n_\mathrm{on}) = 11.4\pm1.0$ MeV and $B_\mathrm{SNM}(n_\mathrm{on}) = -14.1\pm0.1$ by reproducing finite nuclei properties~\cite{Brown2013_PRL111-232502}, where $n_\mathrm{on}$ is approximately the average baryon number density of finite nuclei and the constraint is fulfilled by the predictions of both functionals adopted here. More detailed discussions can be found in Ref.~\cite{Wei2020_CPC44-074107}.

\subsection{\label{sec:EOS_Nonuniform} Nonuniform nuclear matter}

\subsubsection{\label{sec:EOS_pasta} Nuclear pasta at $n_\mathrm{b}\geq 10^{-4}$ fm${}^{-3}$}

The microscopic structures of nuclear matter are obtained by solving the Klein-Gordon equations~(\ref{eq:KG_sigma}-\ref{eq:KG_photon}) in a WS cell with the density distributions of fermions fixed by Eq.~(\ref{eq:chem_cons}). To simplify our calculation, instead of solving Eqs.~(\ref{eq:KG_sigma}-\ref{eq:KG_photon}) and~(\ref{eq:chem_cons}) inside a large 3D periodic cell including exact WS cells~\cite{Okamoto2012_PLB713-284, Okamoto2013_PRC88-025801, Xia2021_PRC103-055812}, we have adopted the spherical and cylindrical approximations~\cite{Maruyama2005_PRC72-015802}. The differential equations for the mean fields ($\phi=\sigma$, $\omega$, $\rho$, $A$) are then reduced to one-dimensional, i.e.,
\begin{eqnarray}
 \mathrm{1D:}\ \ \ \  && \nabla^2 \phi(\vec{r}) = \frac{\mbox{d}^2\phi(r)}{\mbox{d}r^2}; \label{eq:dif_1D} \\
 \mathrm{2D:}\ \ \ \  && \nabla^2 \phi(\vec{r}) = \frac{\mbox{d}^2\phi(r)}{\mbox{d}r^2} + \frac{1}{r} \frac{\mbox{d}\phi(r)}{\mbox{d}r}; \label{eq:dif_2D}\\
 \mathrm{3D:}\ \ \ \  && \nabla^2 \phi(\vec{r}) = \frac{\mbox{d}^2\phi(r)}{\mbox{d}r^2} + \frac{2}{r} \frac{\mbox{d}\phi(r)}{\mbox{d}r}. \label{eq:dif_3D}
\end{eqnarray}
Those differential equations are solved with fast cosine transformation as illustrated in Ref.~\cite{Xia2021_PRC103-055812}, which satisfies the boundary conditions $\left.\frac{\mbox{d}\phi(r)}{\mbox{d}r}\right|_{r=0,R_\mathrm{W}}=0$ with $R_\mathrm{W}$ being the WS cell radius. By fulfilling the constancy of chemical potentials in Eq.~(\ref{eq:chem_cons}), these conditions in fact correspond to the reflective boundary conditions at $r=0$ and $r=R_\mathrm{W}$. The optimum cell size $R_\mathrm{W}$ is then fixed by minimizing the energy per baryon of nuclear matter at fixed average baryon number density $n_\mathrm{b}$ and proton fraction $Y_p$, where the contribution of electrons are included  fulfilling the global charge neutrality condition
\begin{equation}
  \int \left[n_p(\vec{r}) - n_e(\vec{r})\right] \mbox{d}^3 r\equiv 0.
\end{equation}
By assuming various dimensions with geometrical symmetries, five types of pasta phases can be obtained based on TFA, i.e., the slab phase in Eq.~(\ref{eq:dif_1D}),  the rod/tube phases in Eq.~(\ref{eq:dif_2D}), and the droplet/bubble phases in Eq.~(\ref{eq:dif_3D}). The density profiles are then fixed with the constancy of chemical potentials in Eq.~(\ref{eq:chem_cons}).

\begin{figure*}[htbp]
\begin{minipage}[t]{0.33\linewidth}
\centering
\includegraphics[width=\textwidth]{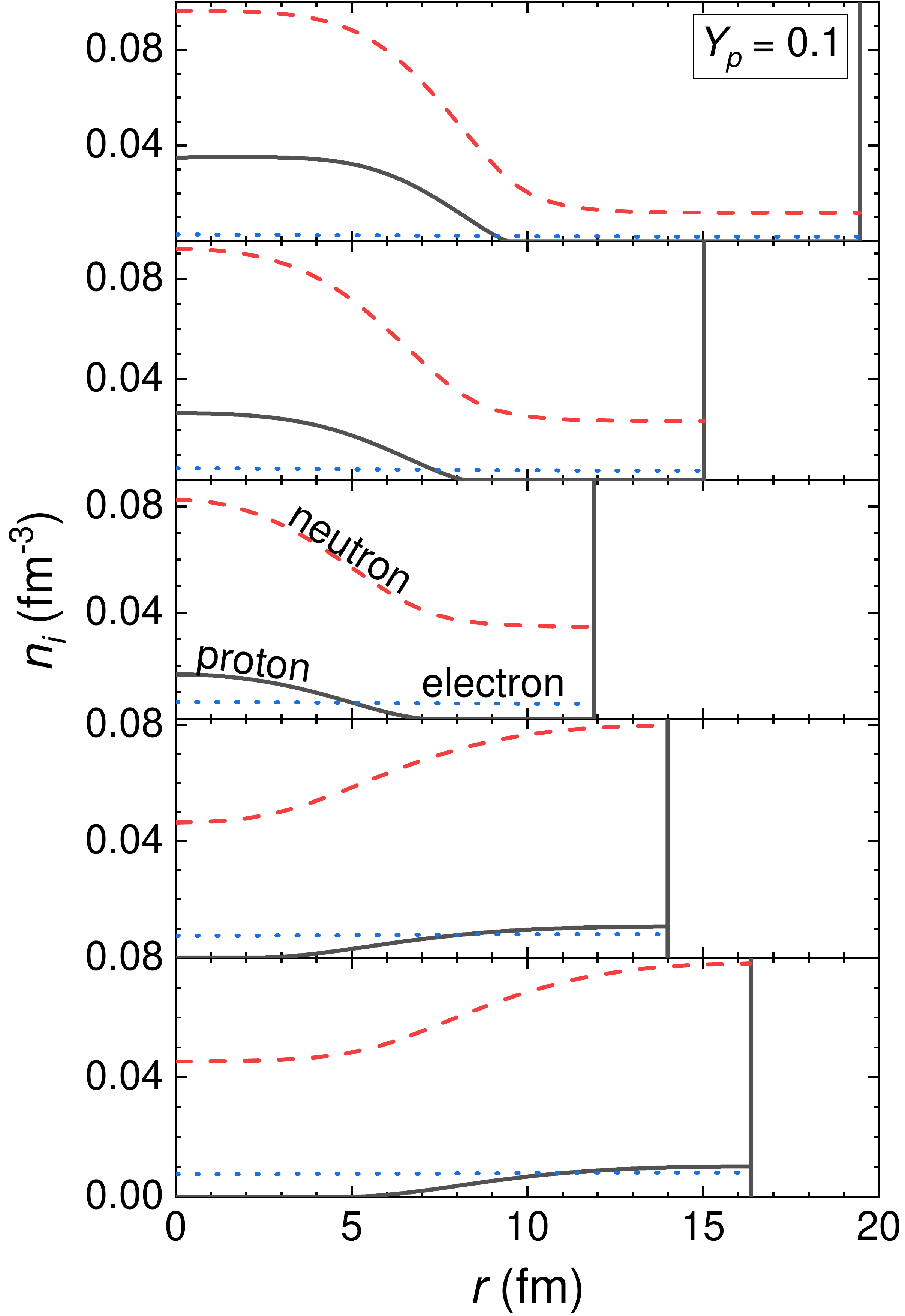}
\end{minipage}%
\hfill
\begin{minipage}[t]{0.32\linewidth}
\centering
\includegraphics[width=\textwidth]{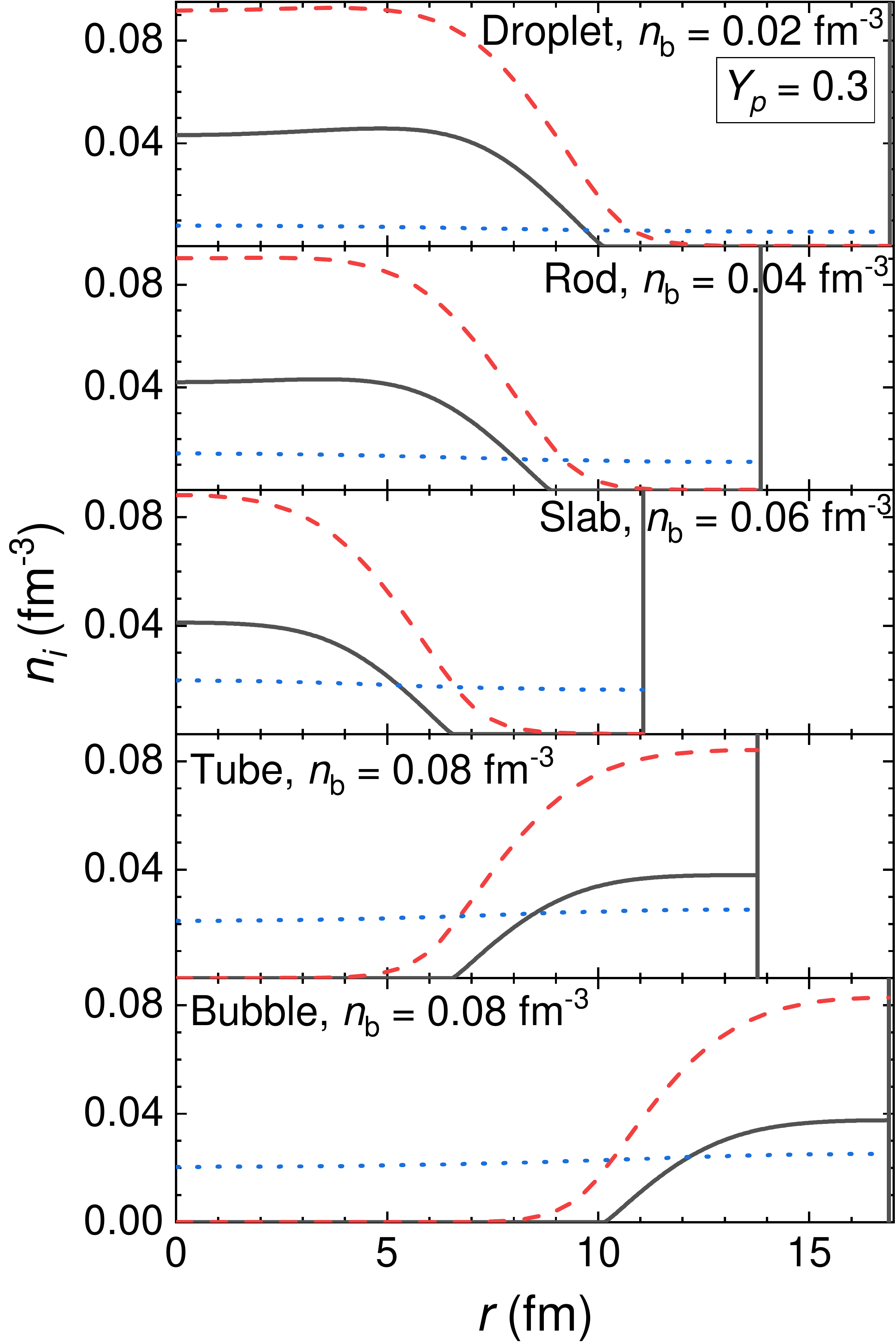}
\end{minipage}
\hfill
\begin{minipage}[t]{0.32\linewidth}
\centering
\includegraphics[width=\textwidth]{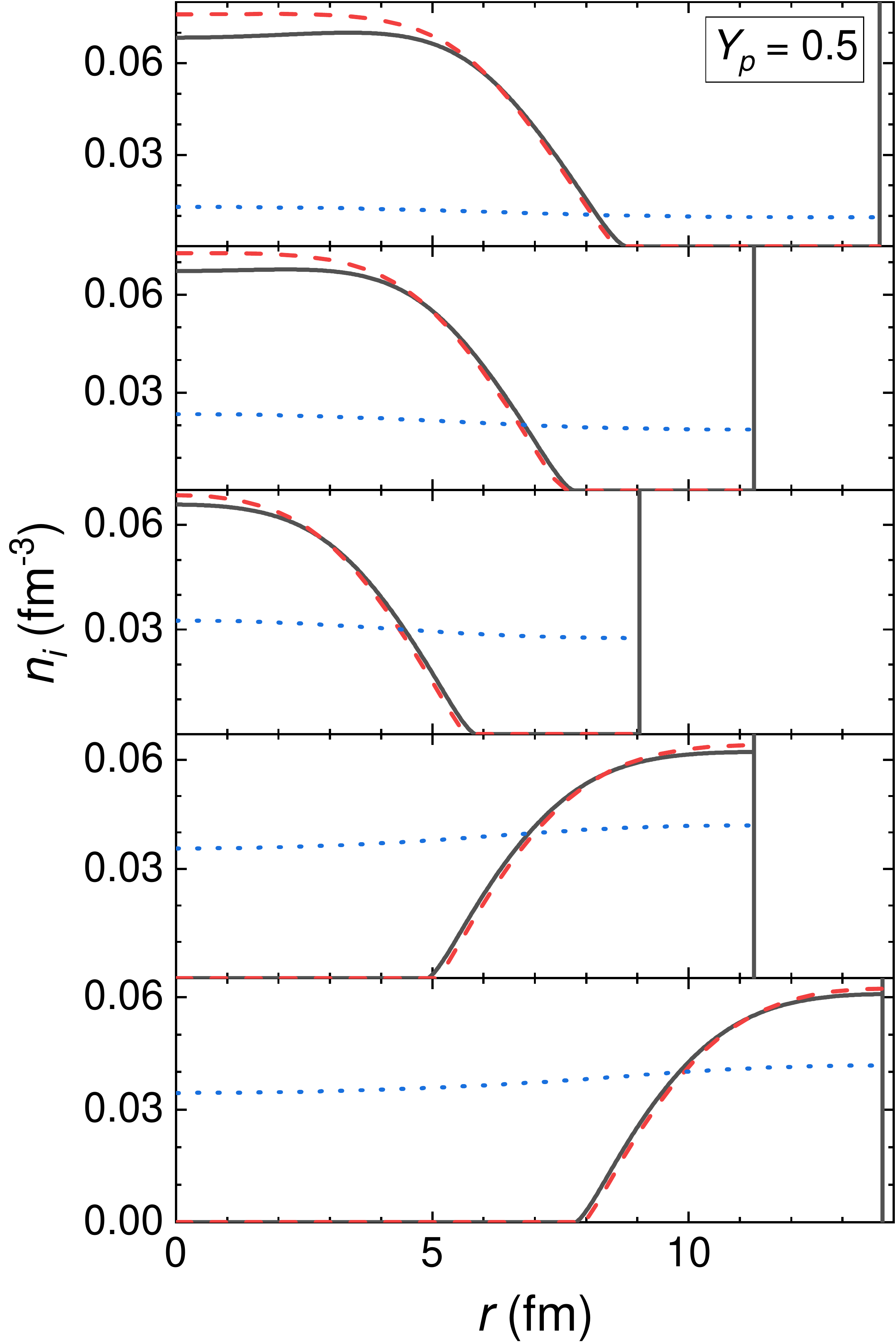}
\end{minipage}
\caption{\label{Fig:Dens_DDLZ1} Density profiles of nucleons and electrons in WS cells for droplet, rod, slab, tube, and bubble phases at $n_\mathrm{b}=0.02$, 0.04, 0.06, 0.08, and 0.08 fm$^{-3}$ (from top to bottom), respectively. Both asymmetric nuclear matter ($Y_p = 0.1$, 0.3) and symmetric nuclear matter ($Y_p = 0.5$) are considered, where the covariant density functional DD-LZ1~\cite{Wei2020_CPC44-074107} is adopted. The boundary of WS cell is indicated by a vertical line in each panel.}
\end{figure*}

Adopting the two covariant density functionals DD-LZ1~\cite{Wei2020_CPC44-074107} and DD-ME2~\cite{Lalazissis2005_PRC71-024312}, the properties of typical pasta structures (droplet, rod, slab, tube, and bubble) are obtained. As an example, in Fig.~\ref{Fig:Dens_DDLZ1} we present the density profiles for typical pasta phases at various densities and proton fractions adopting the functional DD-LZ1. The droplet size $R_\mathrm{d}$ and WS cell size $R_\mathrm{W}$ can be obtained with
\begin{equation}
 R_\mathrm{d} =
 \left\{\begin{array}{l}
   R_\mathrm{W}\left(\frac{\langle n_p \rangle^2}{\langle n_p^2 \rangle}\right)^{1/D},  \text{\ \ \ \ \ \ \  droplet-like}\\
   R_\mathrm{W} \left(1- \frac{\langle n_p \rangle^2}{\langle n_p^2 \rangle}\right)^{1/D},  \text{\ \ bubble-like}\\
 \end{array}\right.,  \label{Eq:Rd}
\end{equation}
where $\langle n_p^2 \rangle = \int n_p^2(\vec{r}) \mbox{d}^3 r/V$ and $\langle n_p \rangle  = \int n_p(\vec{r}) \mbox{d}^3 r/V$ with the WS cell volume
\begin{equation}
  V =
 \left\{\begin{array}{l}
   \frac{4}{3}\pi R_\mathrm{W}^3,\  D = 3\\
   \pi a R_\mathrm{W}^2 , \  D = 2\\
   a^2 R_\mathrm{W}, \ \  D = 1\\
 \end{array}\right.. \label{Eq:V}
\end{equation}
Here $D$ represents the dimension with $D = 3$ for droplets/bubbles, $D = 2$ for rods/tubes, and $D = 1$ for slabs. Since the slab and rod/tube extend infinitely in space for $D = 1$ and 2, we have adopted a finite cell size $a$ so that the volume is finite. Note that the density profiles end at the cell boundary $r=R_\mathrm{W}$, which is indicated by a vertical line in each panel. It is evident that the density distributions of electrons are not constant. This leads to charge screening effects and affects the properties of nuclear pastas~\cite{Maruyama2005_PRC72-015802}, which would become significant at large proton fractions, e.g., $Y_p=0.5$. Neutron starts to drip out and form neutron gas outside of the nucleus as we decrease the proton fraction (e.g., $Y_p = 0.1$), where the neutron density never vanish throughout the WS cell. For protons, on the contrary, the density always drops to zero outside of the nucleus. Meanwhile, comparing the density profiles determined by the two functionals, we find both of which are similar to each other, while those obtained with DD-LZ1 vary more smoothly than that of DD-ME2.

\subsubsection{\label{sec:EOS_outer} Droplet phase at $n_\mathrm{b}<10^{-4}$ fm${}^{-3}$}
For nuclear matter at $n_\mathrm{b}<10^{-4}$ fm${}^{-3}$, we consider only the droplet phase since it is energetically more favorable.
As we decrease the density of nuclear matter, the optimum cell size for the WS cell grows drastically, which quickly exceeds the limit for any viable numerical simulations illustrated in Sec.~\ref{sec:EOS_pasta}. In such cases, we divide a WS cell into two parts, i.e., a core with radius $R_\mathrm{in}$ and a spherical shell ($R_\mathrm{in}<r\leq R_\mathrm{W}$) covering the core. Electrons and neutrons in the shell region take constant densities. To retain the effects of charge screening as much as possible, the electrons still move freely within the core at $r<R_\mathrm{in}$, while the electron and neutron densities in the shell region are fixed by minimizing the energy per baryon at given average baryon number density $n_\mathrm{b}$, core radius $R_\mathrm{in}$, and WS cell size $R_\mathrm{W}$. We note that the optimum density distributions are still consistent with the constancy of chemical potentials in Eq.~(\ref{eq:chem_cons}), where the chemical potentials for each type of particles in the shell region are in fact their average values.

\begin{figure}[!ht]
  \centering
  \includegraphics[width=\linewidth]{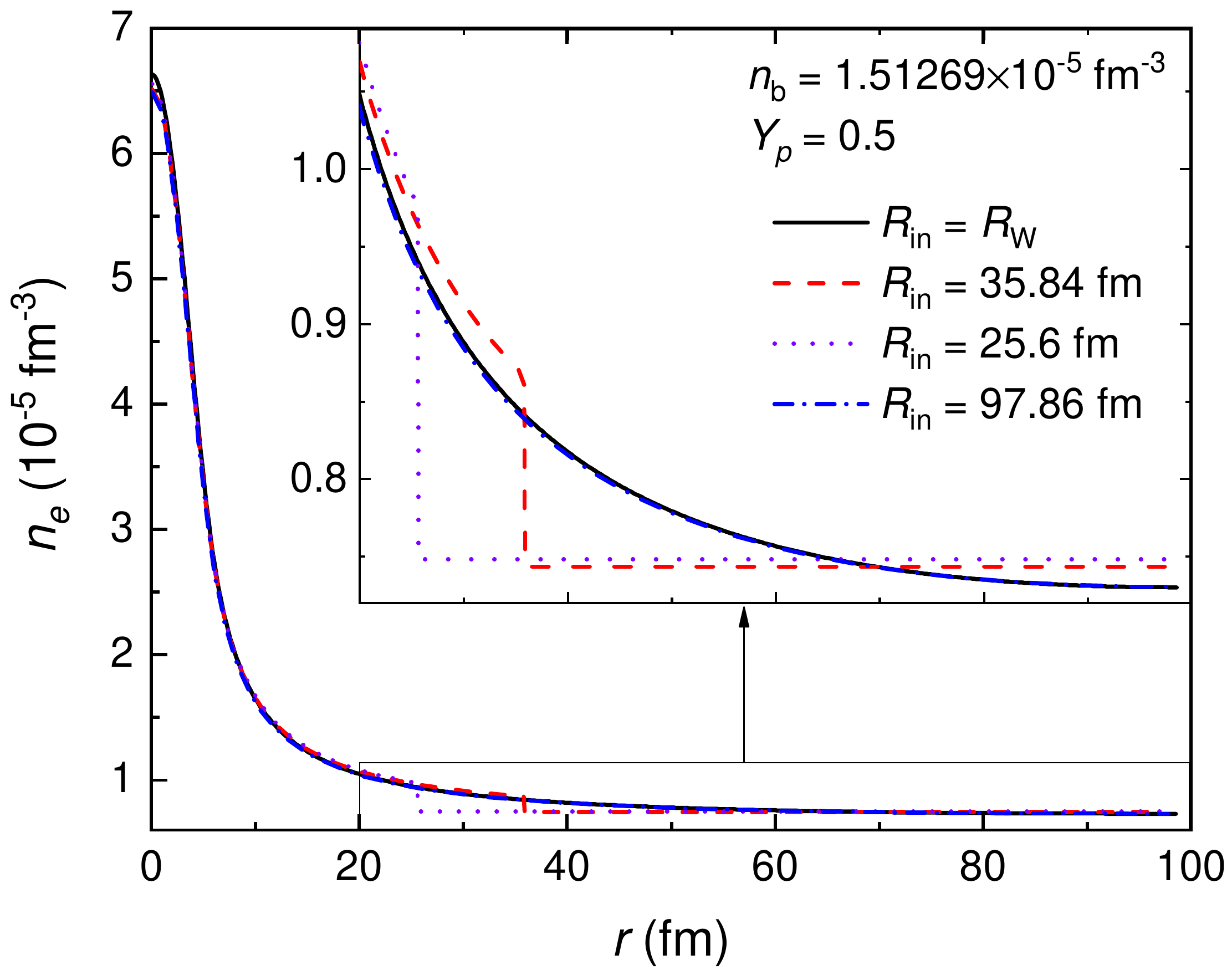}
  \caption{\label{Fig:eDens_aprx}Density profiles of electrons in WS cells for various core radius $R_\mathrm{in}$, in comparison with that of the full calculation ($R_\mathrm{W} = 98.63$ fm) in Sec.~\ref{sec:EOS_pasta}.}
\end{figure}

As an example, in Fig.~\ref{Fig:eDens_aprx} we present the density profiles of electrons in WS cells for the droplet phase of nuclear matter at an average baryon number density $n_\mathrm{b}=1.51269\times10^{-5}$ fm$^{-3}$ and proton fraction $Y_p = 0.5$, where the covariant density functional DD-LZ1~\cite{Wei2020_CPC44-074107} is adopted. The reflective boundary conditions at $r=0$ and $r=R_\mathrm{W}$ are fulfilled with $\left.\frac{\mbox{d}n_e(r)}{\mbox{d}r}\right|_{r=0,R_\mathrm{W}}=0$. A spherical nucleus is located in the center at $r=0$, which attracts electrons so that the densities decrease with $r$. Note that neutrons are still confined within the nucleus and thus vanish in the shell region. The WS cell size $R_\mathrm{W}$ is optimized for each core radius $R_\mathrm{in}$, which is slightly smaller comparing with that of the full calculation with $R_\mathrm{in} =R_\mathrm{W}$. It is evident that the density profiles obtained in each scenario coincide with each other in the core region, while the density in the shell region takes constant values and is sensitive to the choice of the core radius $R_\mathrm{in}$. The corresponding thermodynamic quantities such as the energy density and pressure vary little with respect to $R_\mathrm{in}$, suggesting that the EOSs obtained by dividing the WS cell into two parts are consistent with that of the full calculation.

\begin{figure}[!ht]
  \centering
  \includegraphics[width=\linewidth]{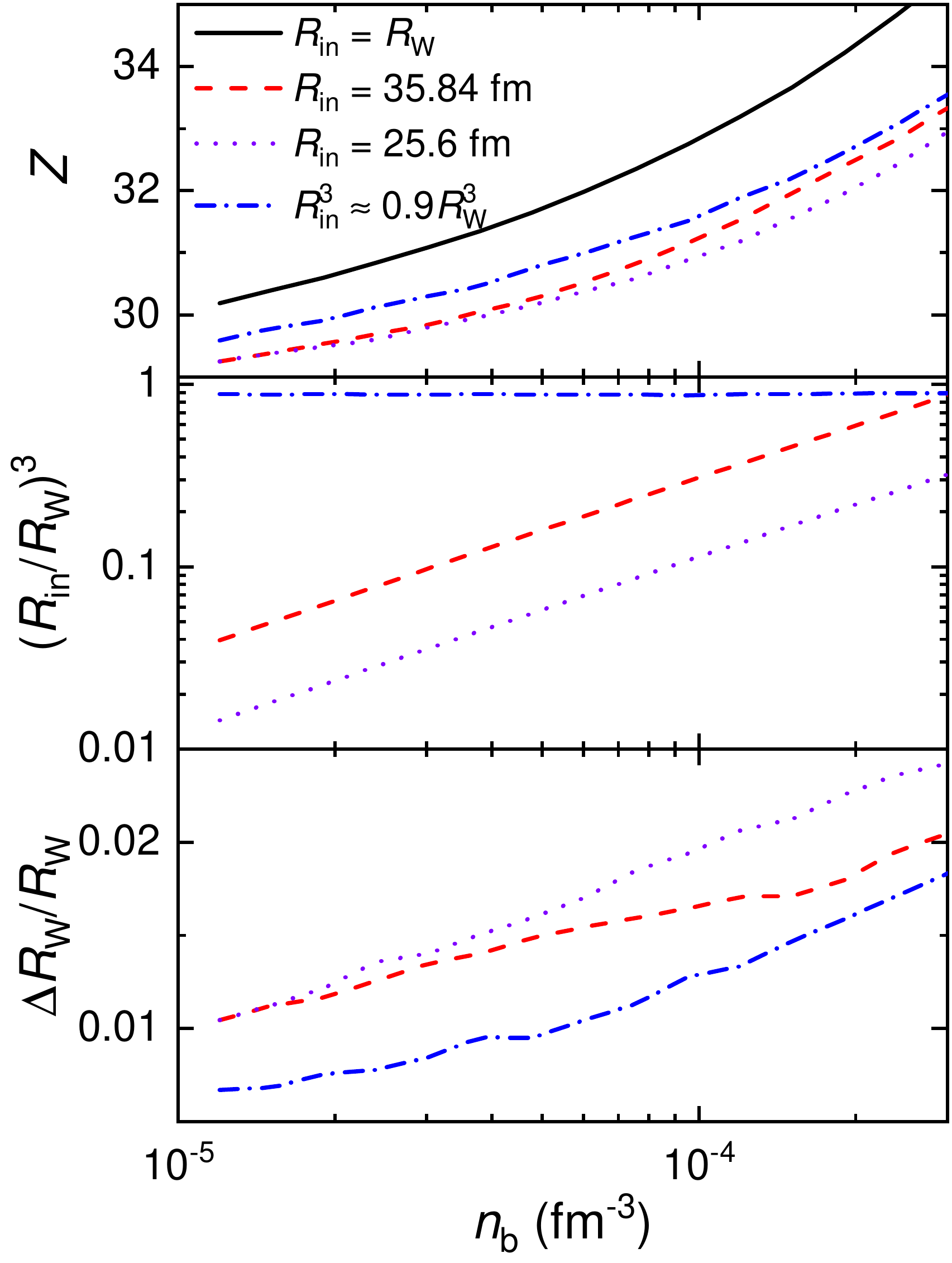}
  \caption{\label{Fig:Charge_aprx} Proton number, volume fraction of the core, relative deviation of the WS cell size for the droplet phase of symmetric nuclear matter ($Y_p = 0.5$) as functions of baryon number density $n_\mathrm{b}$. }
\end{figure}

In contrast to the EOSs, the proton ($Z$) and neutron ($N$) numbers of the nucleus are altered if $R_\mathrm{in} < R_\mathrm{W}$, which is due to the slight variations in the optimum WS cell radius $R_\mathrm{W}$. To show this explicitly, in Fig.~\ref{Fig:Charge_aprx} we present the charge number $Z$, volume fraction of the core to WS cell $R_\mathrm{in}^3/R_\mathrm{W}^3$, relative deviation of the WS cell size $\Delta R_\mathrm{W}/R_\mathrm{W}$ ($\Delta R_\mathrm{W} = \left.R_\mathrm{W}\right|_{R_\mathrm{in}=R_\mathrm{W}} - \left.R_\mathrm{W}\right|_{R_\mathrm{in}<R_\mathrm{W}}$) for the droplet phase of symmetric nuclear matter ($Y_p = 0.5$) as functions of baryon number density $n_\mathrm{b}$. Three different core radii $R_\mathrm{in}$ are adopted, where the corresponding results are compared with those of the full calculation at $R_\mathrm{in}=R_\mathrm{W}$. It is found that dividing the WS cell into two parts leads to an underestimation of the proton/neutron numbers, where the reduction increases if smaller core radius $R_\mathrm{in}$ is adopted. Note that the corresponding optimum WS cell sizes are altered slightly, suggesting that the proton/neutron numbers of the nucleus are sensitive to the microscopic structures of WS cell.  Nevertheless, it is worth mentioning that instead of solving the Dirac equations, we have adopted TFA for nuclear pastas, which introduces uncertainty for the proton/neutron numbers as the shell effects are not accounted for in our calculation. At smaller densities, the volume fraction of the core decreases quickly, while the deviation of proton/neutron numbers from full calculation decreases. In such cases, to obtained the EOSs of nuclear matter at densities $n_\mathrm{b}\lesssim 10^{-4}$ fm${}^{-3}$, we take a moderate value with $R_\mathrm{in} = 35.84$ fm.

\section{\label{sec:results} Results and Discussion}
\begin{figure}[!ht]
  \centering
  \includegraphics[width=\linewidth]{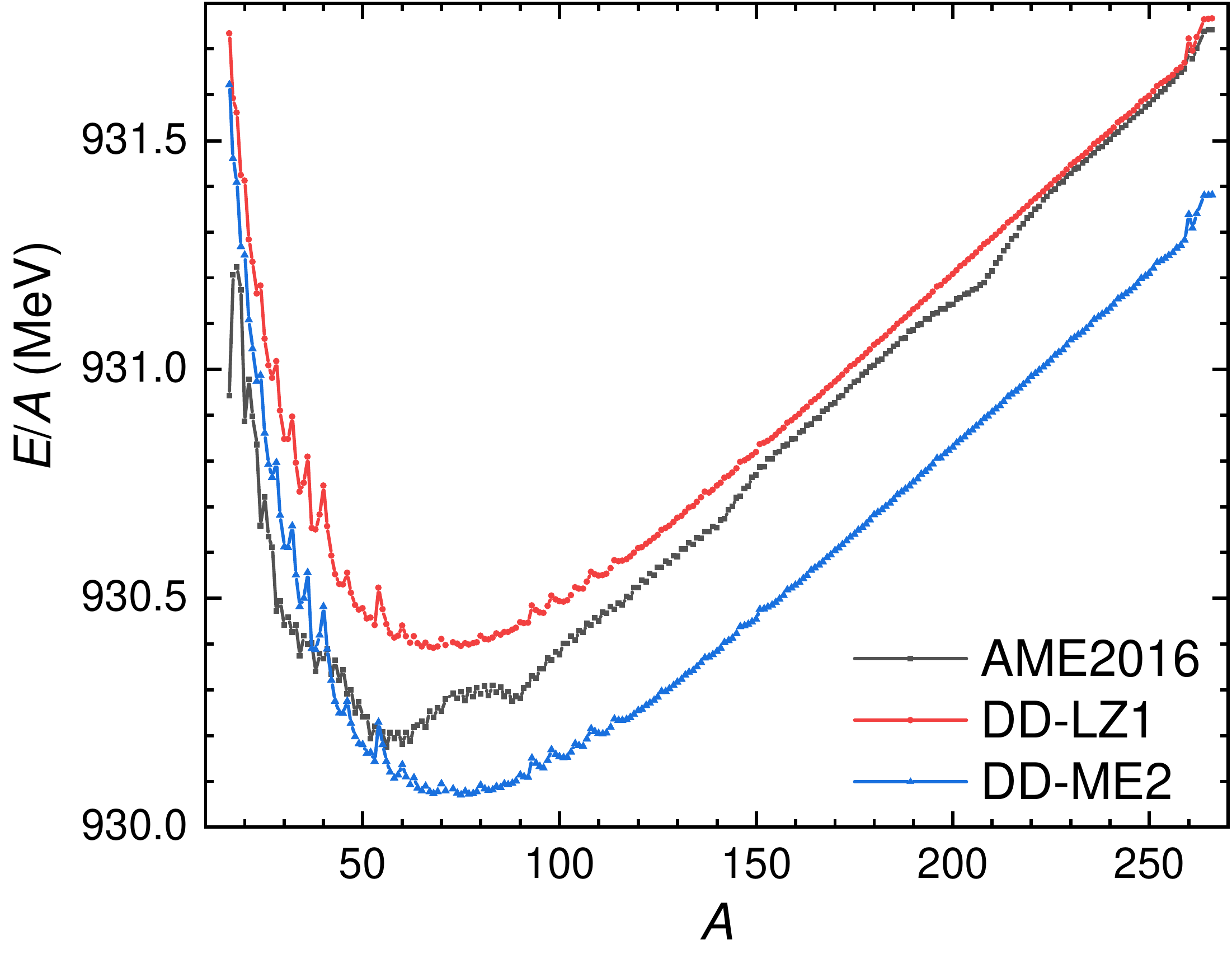}
  \caption{\label{Fig:EpA_Nucl}Energy per baryon of finite nuclei obtained with the two covariant density functionals DD-LZ1~\cite{Wei2020_CPC44-074107} and DD-ME2~\cite{Lalazissis2005_PRC71-024312} in the framework of TFA. The results are compared with the experimental data from AME2016~\cite{Audi2017_CPC41-030001, Huang2017_CPC41-030002, Wang2017_CPC41-030003}.}
\end{figure}

To give a rough estimation on the uncertainties of our calculation, in Fig.~\ref{Fig:EpA_Nucl} we first present the energy per baryon of finite nuclei obtained within the framework of TFA. Note that the spikes in Fig.~\ref{Fig:EpA_Nucl} emerge due to the variations in the energy per baryon along the isotopic chain. The results are then compared with the experimental value from the 2016 Atomic Mass Evaluation (AME2016)~\cite{Audi2017_CPC41-030001, Huang2017_CPC41-030002, Wang2017_CPC41-030003}. Since the shell effects and nucleon pairing~\cite{Furtado2021_JPG49-025202} are neglected in our current study, the obtained results deviate from the experimental value, which is particularly the case for nuclei with proton/neutron numbers close to the magic numbers. In the framework of TFA, it is found that DD-LZ1 gives a better description for heavy nuclei, while the energy per baryon of light nuclei obtained with DD-ME2 are close to the experimental value. In such cases, we expect that the energy density of nuclear pasta obtained by DD-LZ1 is more accurate as the nuclei are heavier. Nevertheless, it is worth mentioning that there is a systematic underestimation for DD-ME2 on the energy per baryon, which is mainly due to the smaller nucleon mass adopted in the calculation with $M=938.5$ MeV and 938.9 MeV for DD-ME2 and DD-LZ1, respectively. If we compare the binding energies of nuclei, the results predicted by the two functionals coincide with each other and converge for heavy nuclei at large mass numbers.

\subsection{\label{sec:pasta_fix} Nuclear matter EOSs at fixed proton fractions}

\begin{figure*}[htbp]
\begin{minipage}[t]{0.33\linewidth}
\centering
\includegraphics[width=\textwidth]{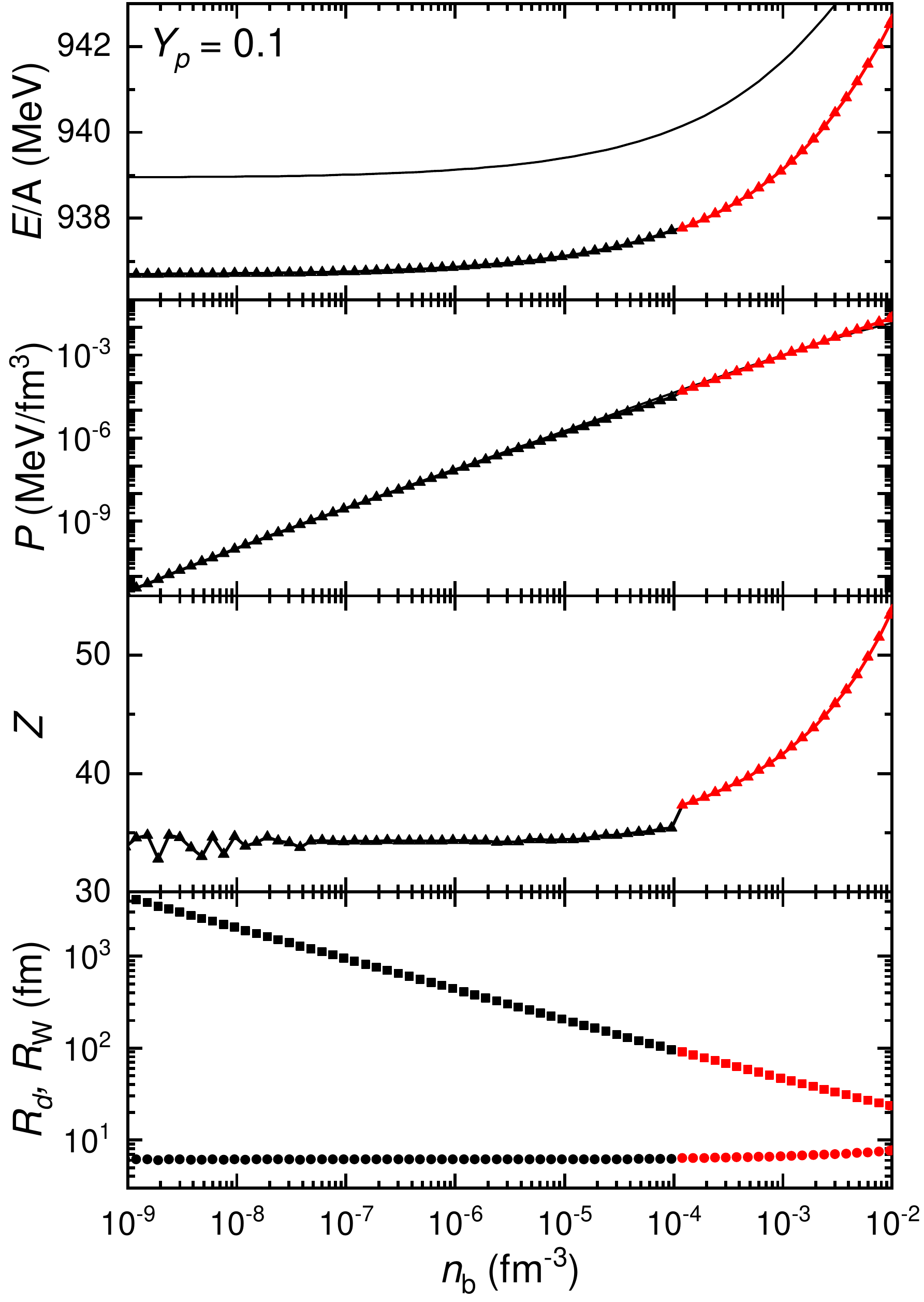}
\end{minipage}%
\hfill
\begin{minipage}[t]{0.33\linewidth}
\centering
\includegraphics[width=\textwidth]{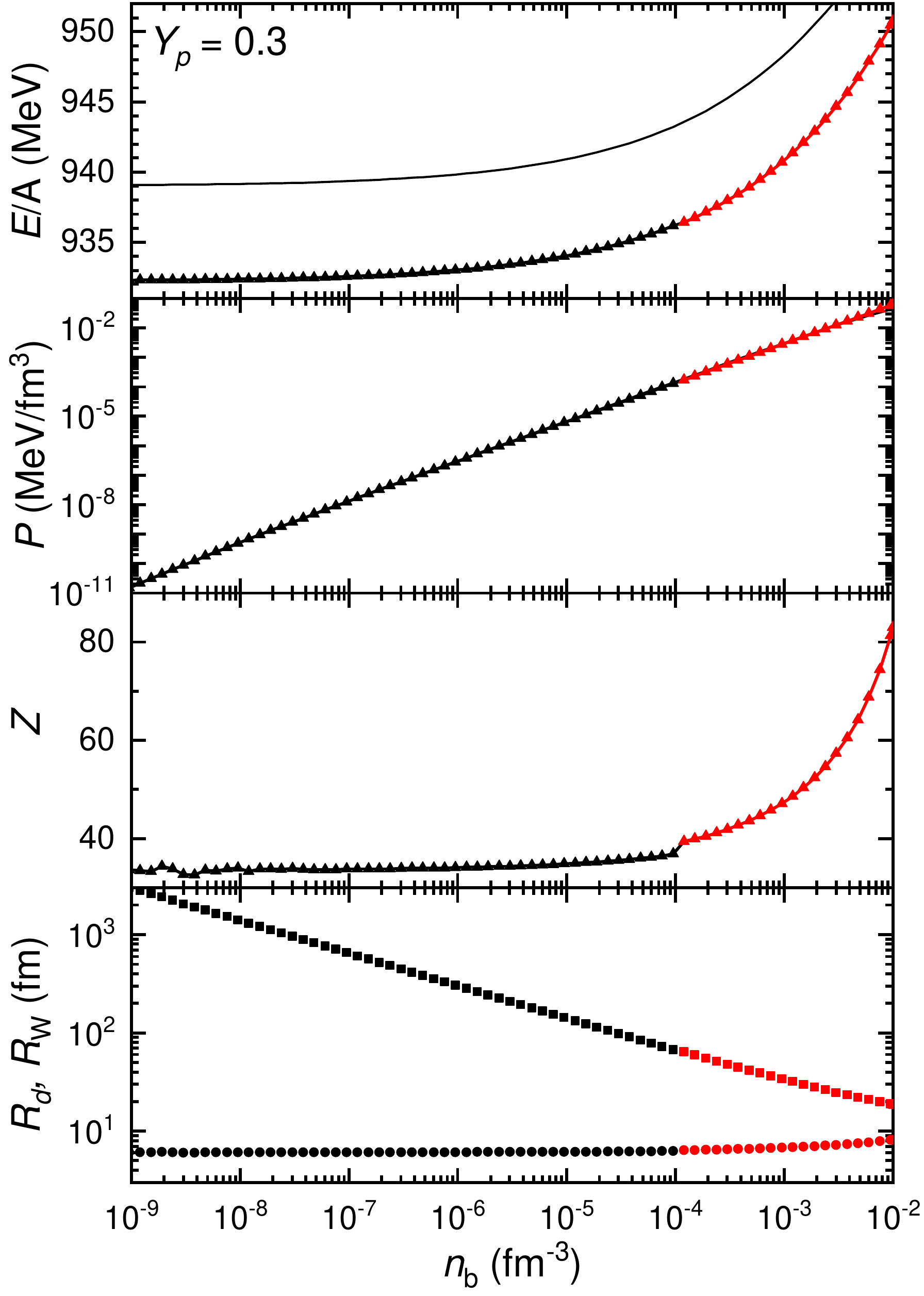}
\end{minipage}
\hfill
\begin{minipage}[t]{0.33\linewidth}
\centering
\includegraphics[width=\textwidth]{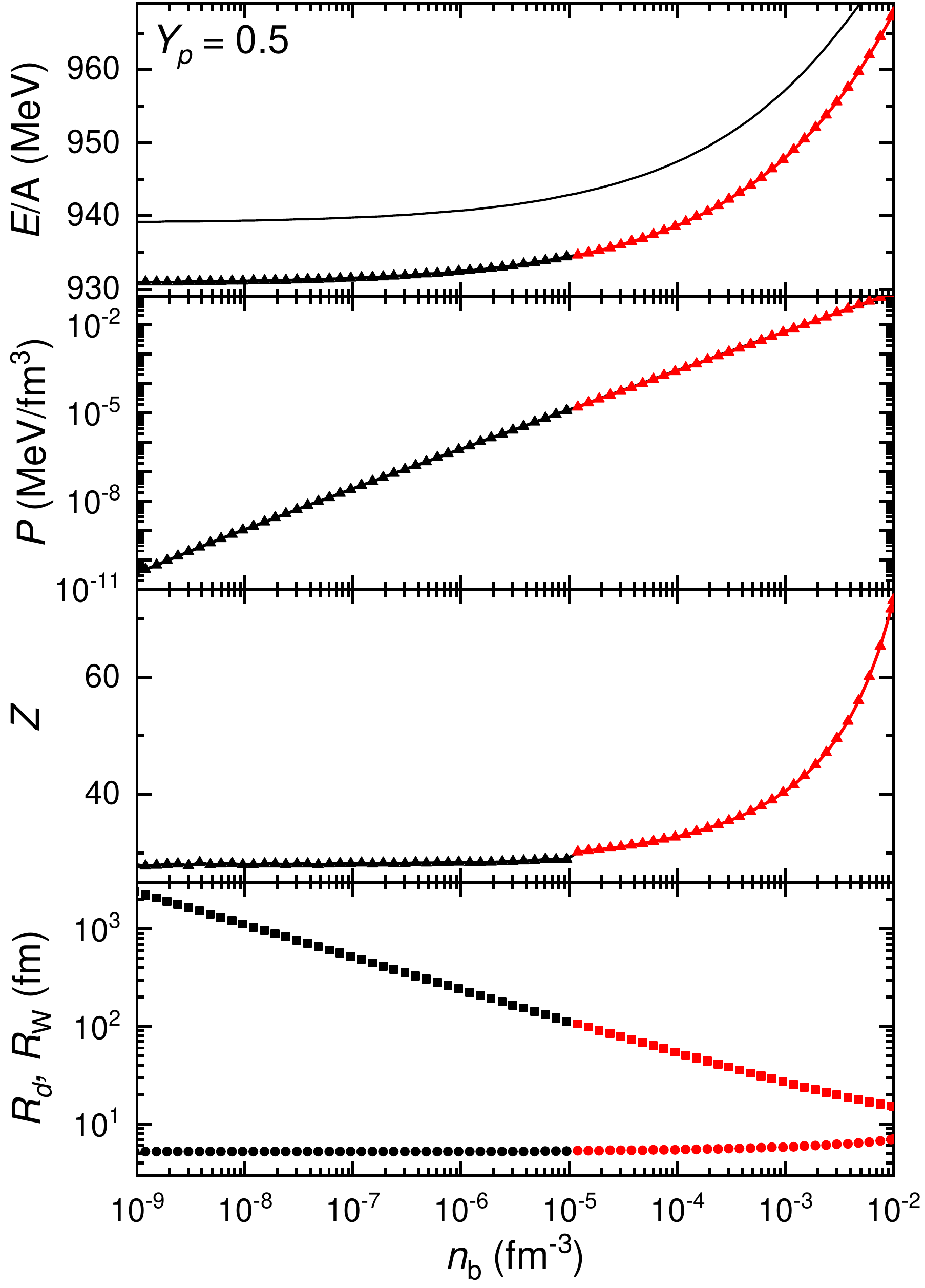}
\end{minipage}
\\
\begin{minipage}[t]{0.31\linewidth}
\centering
\includegraphics[width=\textwidth]{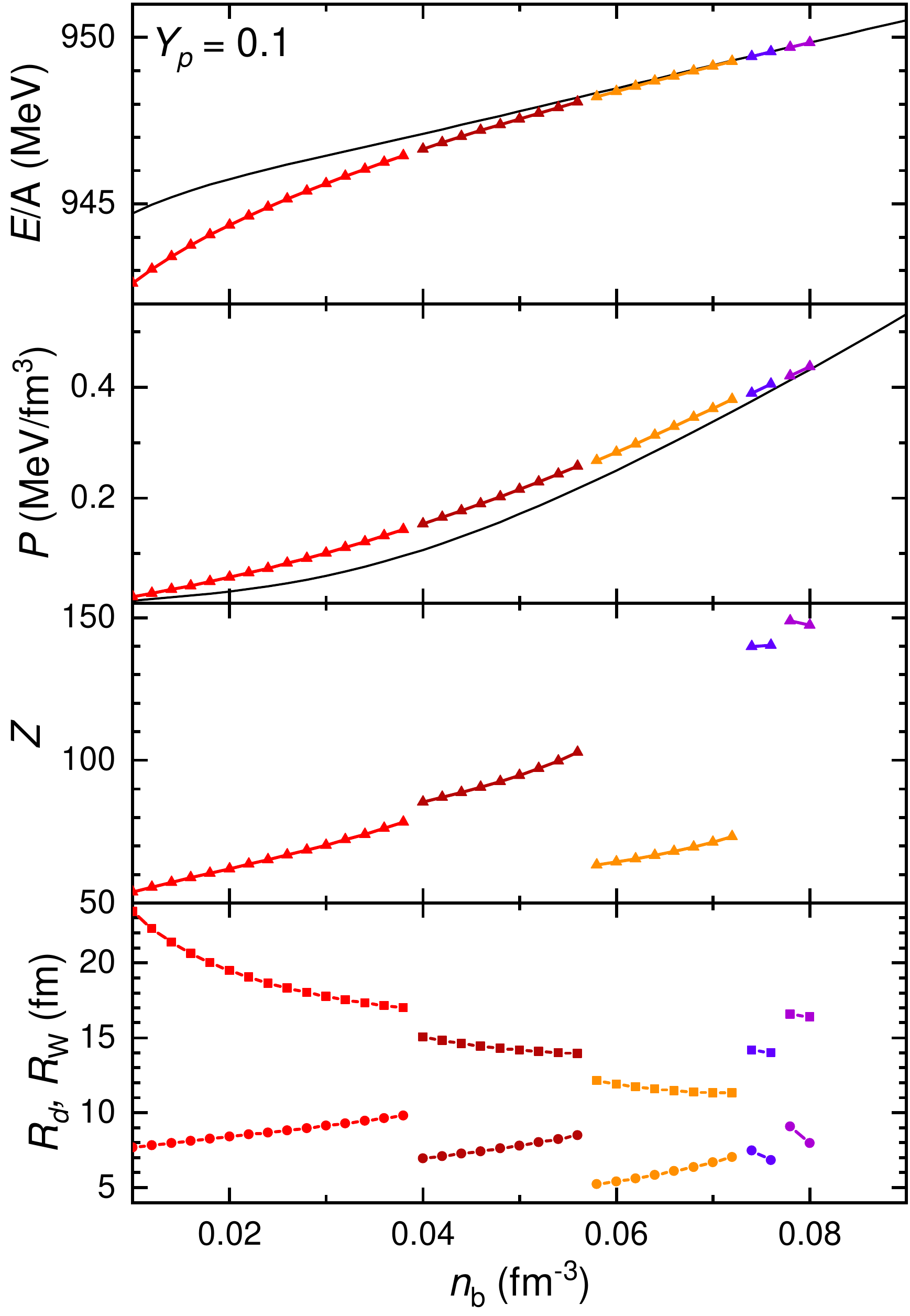}
\end{minipage}%
\hfill
\begin{minipage}[t]{0.31\linewidth}
\centering
\includegraphics[width=\textwidth]{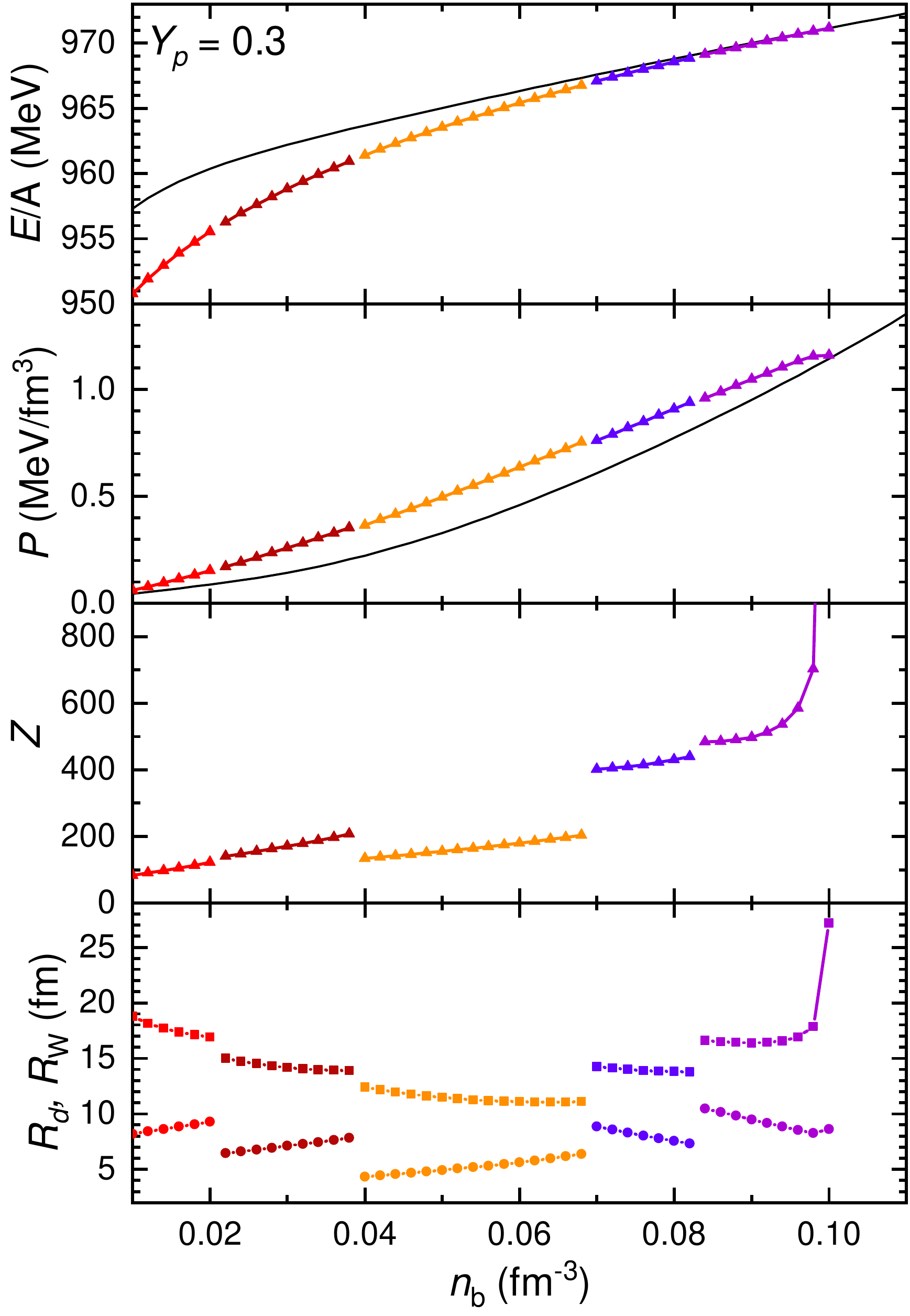}
\end{minipage}
\hfill
\begin{minipage}[t]{0.322\linewidth}
\centering
\includegraphics[width=\textwidth]{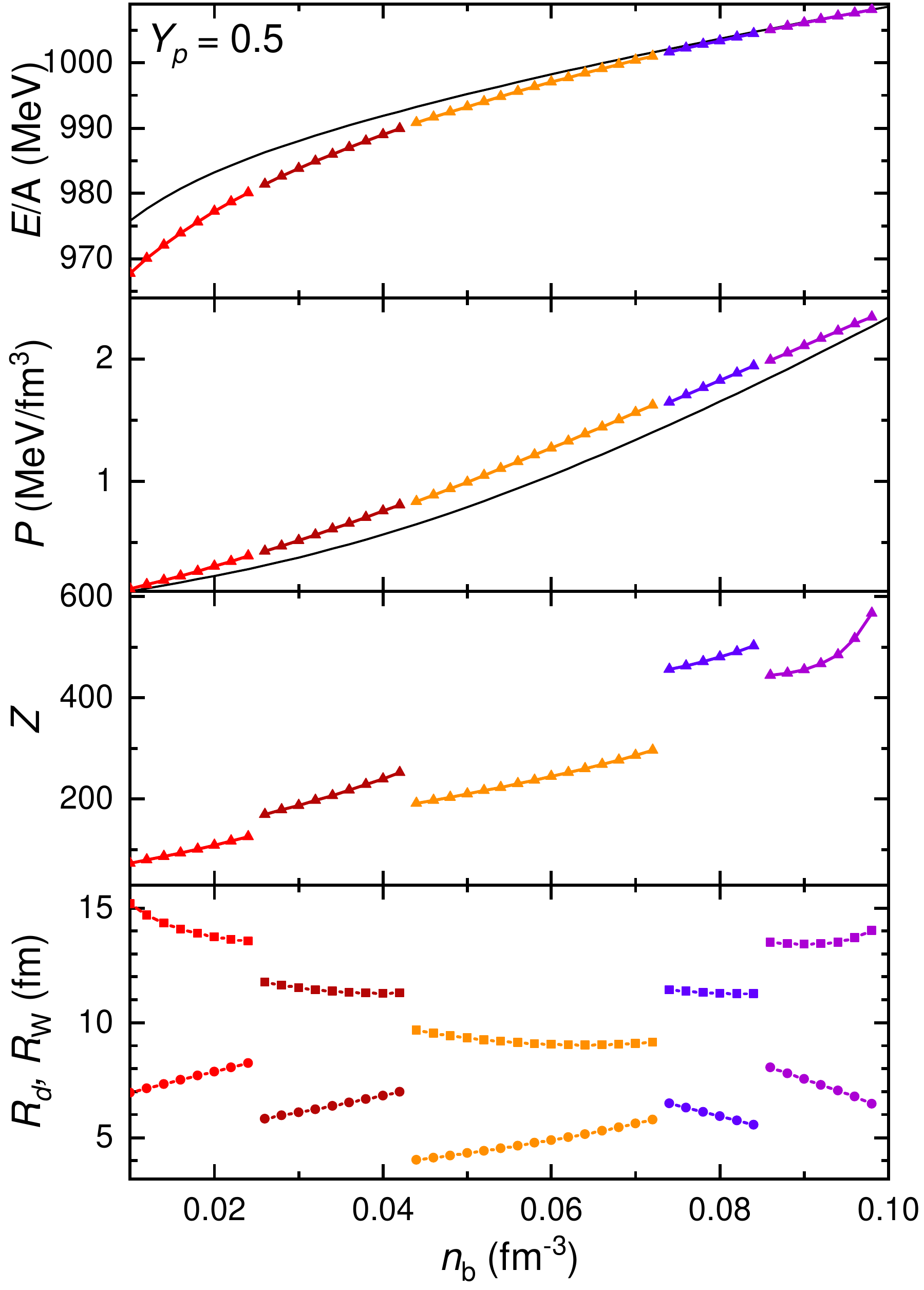}
\end{minipage}
\caption{\label{Fig:EPA_DDLZ1_fix} Energy per baryon ($E/A$), pressure $P$, proton number $Z$ and size $R_\mathrm{d}$ of the droplet, and WS cell size $R_\mathrm{W}$ for asymmetric nuclear matter ($Y_p = 0.1$, 0.3) and symmetric nuclear matter ($Y_p = 0.5$), which are obtained adopting the covariant density functional DD-LZ1~\cite{Wei2020_CPC44-074107} in the framework of TFA. The solid curves indicate the results for uniform matter. The upper panels show the results for the droplet phase at $n_\mathrm{b}<0.01$ fm${}^{-3}$, where the black dots are obtained by dividing a WS cell into two parts as indicated in Sec.~\ref{sec:EOS_outer}. In the lower panels, the droplet (red), rod (dark red), slab (yellow), tube (lavender), and bubble (purple) phases appear sequentially as the density increases.}
\end{figure*}

We first investigate the properties of nuclear matter at fixed proton fractions, i.e., symmetric nuclear matter with $Y_p =0.5$, and asymmetric nuclear matter with $Y_p =0.3$ and 0.1. The baryon number density $n_\mathrm{b}$ ranges from $\sim$$10^{-9}$ to $\sim$0.11 fm${}^{-3}$, where both the uniform and nonuniform phases illustrated in Sec.~\ref{sec:the_EOS} are examined. For nonuniform nuclear matter, the global charge neutrality condition is fulfilled by including the contributions of electrons, which are distributed non-uniformly inside WS cells. The optimum structures of nuclear matter are then fixed by searching for the energy minimum among various shapes and WS cell sizes at fixed baryon number density $n_\mathrm{b}$ and proton fraction $Y_p$.

The energy per baryon ($E/A$), pressure $P$, proton number of nucleus $Z$, droplet size $R_\mathrm{d}$, and WS cell size $R_\mathrm{W}$ for asymmetric nuclear matter ($Y_p = 0.1$, 0.3) and symmetric nuclear matter ($Y_p = 0.5$) are presented in Fig.~\ref{Fig:EPA_DDLZ1_fix}, where the covariant density functionals DD-LZ1~\cite{Wei2020_CPC44-074107} is adopted. The thin solid curves indicate the results for uniform nuclear matter, where the energy per baryon is effectively reduced by up to 10 MeV as nonuniform structures take place. At $n_\mathrm{b} \gtrsim 0.01$ fm${}^{-3}$, the obtained pressure becomes larger for nuclear pastas, while at lower densities the pressure becomes indistinguishable between the uniform and nonuniform phases. In general, the obtained results of DD-LZ1~\cite{Wei2020_CPC44-074107} coincide with those of DD-ME2~\cite{Lalazissis2005_PRC71-024312} due to the similar behavior of the binding energy in Fig.~\ref{Fig:Binding}. Meanwhile, the energy per baryon for DD-ME2 is slightly smaller than that of DD-LZ1, which is attributed to the smaller nucleon mass adopted in the calculation as in Fig.~\ref{Fig:EpA_Nucl}.

\begin{table*}
\caption{\label{table:phase} Densities (in fm${}^{-3}$) for shape transitions, which are obtained by varying the density in a step of 0.002 fm${}^{-3}$. }
\begin{tabular}{c|cccc|cccc} \hline \hline
 Transition        & \multicolumn{4}{c|}{DD-LZ1}                           & \multicolumn{4}{c}{DD-ME2} \\
                   & $Y_p=0.1$ & $Y_p=0.3$ & $Y_p=0.5$ &   $\beta$-stable & $Y_p=0.1$ & $Y_p=0.3$ & $Y_p=0.5$ &   $\beta$-stable   \\ \hline
droplet-rod        &  0.039    & 0.021     &  0.025    &    0.059         &  0.039    &    0.021  & 0.025     & 0.063  \\
rod-slab           &  0.057    & 0.039     &  0.043    &    0.065         &  0.059    &    0.039  & 0.043     & 0.071  \\
slab-tube          &  0.073    & 0.069     &  0.073    &    0.069         &  0.077    &    0.071  & 0.075     & 0.073  \\
tube-bubble        &  0.077    & 0.083     &  0.085    &      -           &  0.081    &    0.085  & 0.091     &   -    \\
tube/bubble-uniform &  0.081    & 0.101     &  0.099   &    0.071         &  0.085    &    0.107  & 0.109     & 0.075  \\ \hline
\end{tabular}
\end{table*}

For the microscopic structures of nuclear matter, few examples concerning the detailed density distributions of various nuclear pasta structures are illustrated in Figs.~\ref{Fig:Dens_DDLZ1} and \ref{Fig:eDens_aprx}. The evolution of the shapes and sizes of nuclear pastas with respect to density can be found in Fig.~\ref{Fig:EPA_DDLZ1_fix}. It is found that the droplet, rod, slab, tube, and bubble phases appear sequentially as density increases, which are marked in various colors with different values of $Z$, $R_\mathrm{d}$, and $R_\mathrm{W}$. The corresponding transition densities among various nonuniform phases are indicated in Tab.~\ref{table:phase}. We note that the core-crust transition density generally increases with proton fraction $Y_p$. Meanwhile, the phase diagrams as well as the transition densities obtained with the two functionals are close to each other, while the density regions for nonuniform structures are slightly larger for DD-ME2 than that of DD-LZ1.

\begin{figure*}
  \centering
  \includegraphics[width=0.6\linewidth]{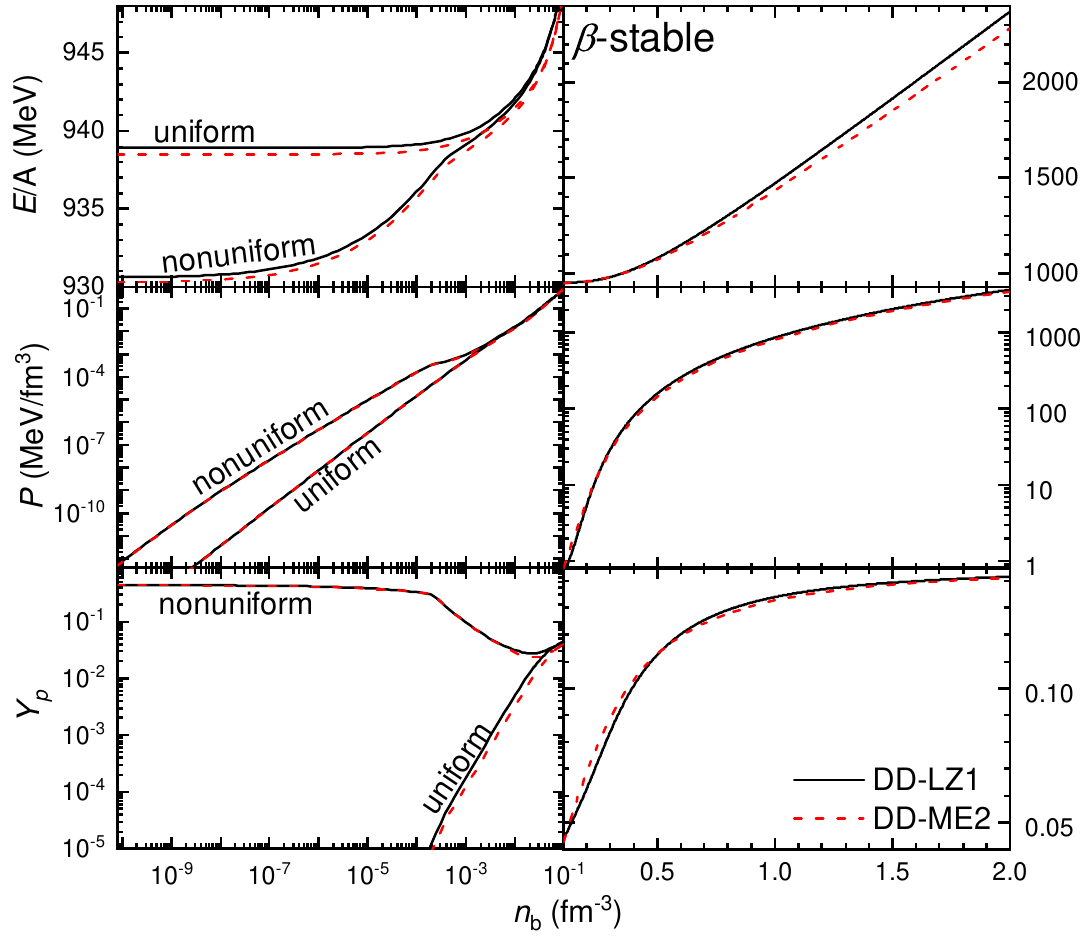}
  \caption{\label{Fig:EOS-beta} The EOSs of cold neutron star matter fulfilling $\beta$-stability condition, which are obtained with the two covariant density functionals DD-LZ1~\cite{Wei2020_CPC44-074107} and DD-ME2~\cite{Lalazissis2005_PRC71-024312}. In the left panels, both the uniform and nonuniform phases are presented, while only uniform phases emerge in the density range of right panels. The corresponding proton factions $Y_p$ are indicated in the bottom panels as well.}
\end{figure*}

The droplet size $R_\mathrm{d}$ and WS cell size $R_\mathrm{W}$ are obtained with Eqs.~(\ref{Eq:Rd}) and (\ref{Eq:V}). According to Fig.~\ref{Fig:EPA_DDLZ1_fix}, it is found that both $R_\mathrm{d}$ and $R_\mathrm{W}$ increase with $D$, which coincide with previous studies~\cite{Maruyama2005_PRC72-015802}. For each configuration, the optimum sizes $R_\mathrm{d}$ and $R_\mathrm{W}$ increase if we adopt smaller proton fraction $Y_p$. Meanwhile, as we decrease the density, the droplet size remains almost constant with $R_\mathrm{d}\approx 6$ fm, while the WS cell size $R_\mathrm{W}$ grows drastically. We thus divide the WS cell into a core with radius $R_\mathrm{in}=35.84$ fm and a spherical shell as illustrated in Sec.~\ref{sec:EOS_outer}.
Comparing the results predicted by the two functionals, aside from the differences in the phase diagrams, we find the obtained droplet sizes $R_\mathrm{d}$ are similar. Nevertheless, the WS cell size $R_\mathrm{W}$ become slightly larger if DD-ME2 is adopted. This leads to slight larger values in the proton ($Z$) and neutron ($N$) numbers for each WS cell, which are determined by $Z=Y_pn_\mathrm{b}V$ and $N=(1-Y_p)n_\mathrm{b}V$ with the volume $V$ fixed by Eq.~(\ref{Eq:V}) with $a = 30$ fm.

\subsection{\label{sec:pasta_beta} Neutron star EOSs in $\beta$-equilibrium}

Now we consider the EOSs of neutron star matter fulfilling the $\beta$-stability condition $\mu_n=\mu_p+\mu_e=\mu_p+\mu_\mu$, where the energy per baryon, pressure, and proton fraction for the most favorable configurations are presented in Fig.~\ref{Fig:EOS-beta}. The obtained results are then compared with that of the uniform matter. With the emergence of nonuniform structures, the proton fractions are increased significantly, which effectively reduces the energy per baryon by up to 8 MeV. The pressure of nonuniform matter becomes larger than that of the uniform one. Comparing the results obtained with the two functionals, we note that the EOSs at densities $n_\mathrm{b} \lesssim 0.01$ fm${}^{-3}$ coincide with each other, while the energy per baryon and consequently the energy density obtained with DD-ME2 is slightly smaller than that of DD-LZ1 (within 0.1\%) due to the smaller nucleon mass adopted in the calculation. By decreasing the density, the energy per baryon decreases and approaches to $\sim$930 MeV, which coincide with the energy per baryon of the most stable nucleus $^{56}$Fe. Compared with previous studies on the EOSs of outer crusts~\cite{Baym1971_ApJ170-299, Haensel1994_AA283-313, Ruester2006_PRC73-035804}, it is found that the differences are insignificant as long as neutrons do not drip out of nuclei, which would effectively soften the EOSs. We note that the slope of the energy per baryon, pressure, and proton fraction change suddenly at $n_\mathrm{b}\gtrsim 2\times 10^{-4}$ fm${}^{-3}$ with $\mu_n>M$, corresponding to the neutron drip density with neutron gas coexists with the liquid phase of nuclear matter.

At vanishing densities, it is found that the proton fraction $Y_p$ approaches to a value slightly smaller than 0.5 in contrast to the cases neglecting Coulomb interaction, where symmetric nuclear matter with $Y_p=0.5$ is more stable. The EOS predicted by DD-LZ1 is softer than that of DD-ME2 at $n_\mathrm{b}\lesssim 0.3$ fm${}^{-3}$, which becomes stiffer at larger densities. Aside from the incompressibility of symmetric nuclear matter, we note that the stiffness of the EOSs is also closely related to the evolution of proton fractions, where the density dependence of symmetry energy play important roles, i.e., $Y_p$ increases with the symmetry energy $S$ and approaches to 0.5. According to Fig.~\ref{Fig:Binding}, the symmetry energy predicted by the functional DD-ME2 is greater than that of DD-LZ1 at $n_\mathrm{b}\lesssim 0.6$ fm${}^{-3}$ but becomes smaller at larger densities. This indicates a smaller curvature parameter $K_\mathrm{sym}$ of symmetric energy for DD-ME2 despite the larger slope $L$, which is attributed to peculiar density dependent behavior of the coupling strengths adopted by DD-LZ1 as indicated in Fig.~\ref{Fig:Coupling}. Note that the stiffness of neutron star matter presented in Fig.~\ref{Fig:EOS-beta} have direct consequences on the mass-radius relations of neutron stars~\cite{Lattimer2012_ARNPS62-485}, which will be illustrated in Sec.~\ref{sec:star}.

\begin{figure}[!ht]
  \centering
  \includegraphics[width=\linewidth]{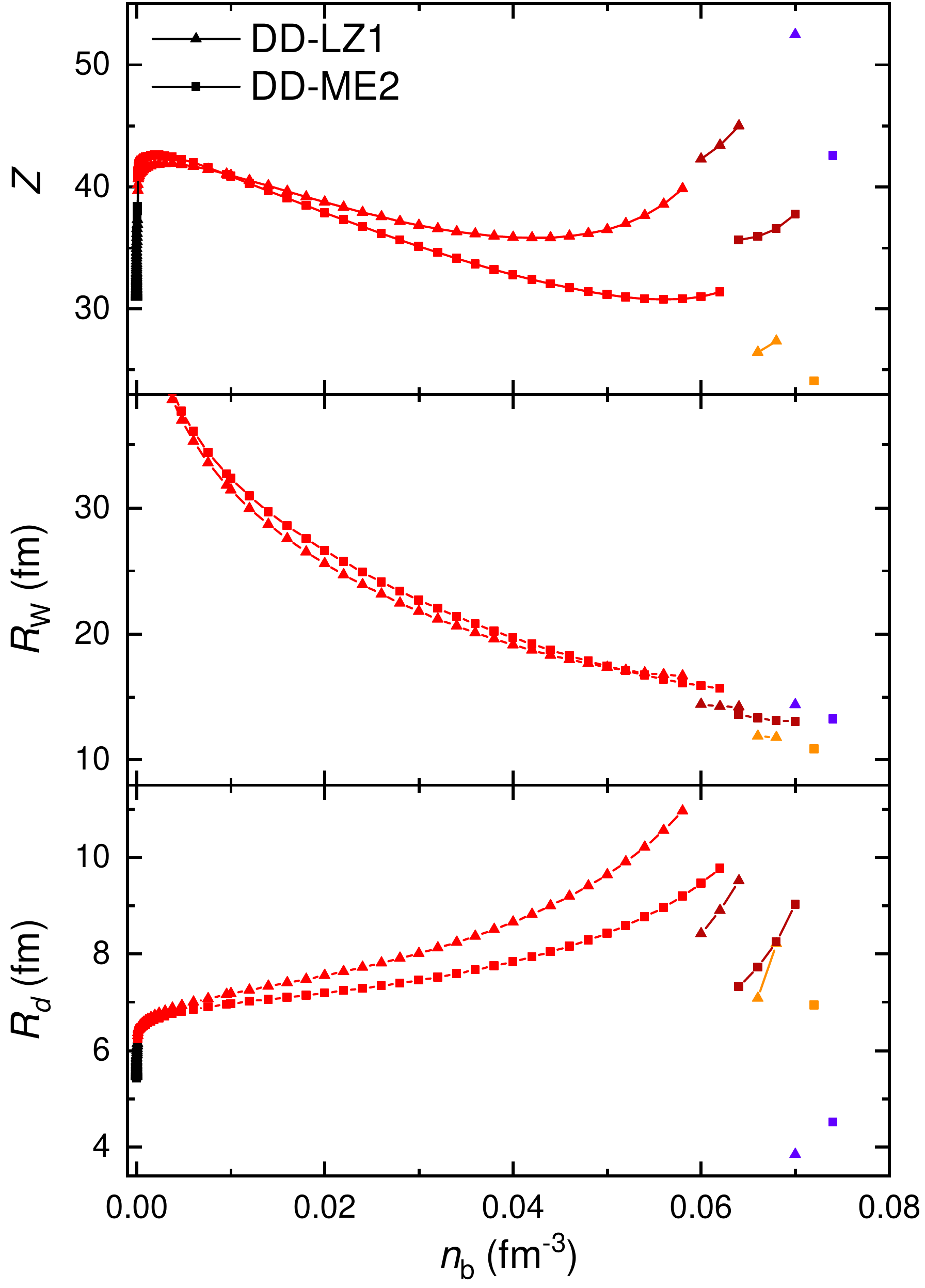}
  \caption{\label{Fig:Micro-beta}Proton number $Z$, WS cell radius $R_\mathrm{W}$, and droplet size $R_\mathrm{d}$ of nuclear pastas corresponding to Fig.~\ref{Fig:EOS-beta}.}
\end{figure}

The microscopic structures of nuclear pasta corresponding to the EOSs in Fig.~\ref{Fig:EOS-beta} are indicated in Fig.~\ref{Fig:Micro-beta}, where the proton number $Z$, WS cell radius $R_\mathrm{W}$, and droplet size $R_\mathrm{d}$ as functions of baryon number density are presented. For the phase diagrams of nuclear pasta in $\beta$-equilibrium, the droplet, rod, slab, tube, and uniform phases appear sequentially as density increases, while the bubble phase does not appear with the energy per baryon being slightly larger ($\sim$0.1 keV). The obtained results with the functional DD-ME2 coincide with those in Ref.~\cite{Grill2012_PRC85-055808}. Nevertheless, as indicated in Tab.~\ref{table:phase}, there are slight differences in the shape transition densities and the emergence of tube phase, which slightly increases the core-crust transition density in our current study.

\begin{figure}[!ht]
  \centering
  \includegraphics[width=\linewidth]{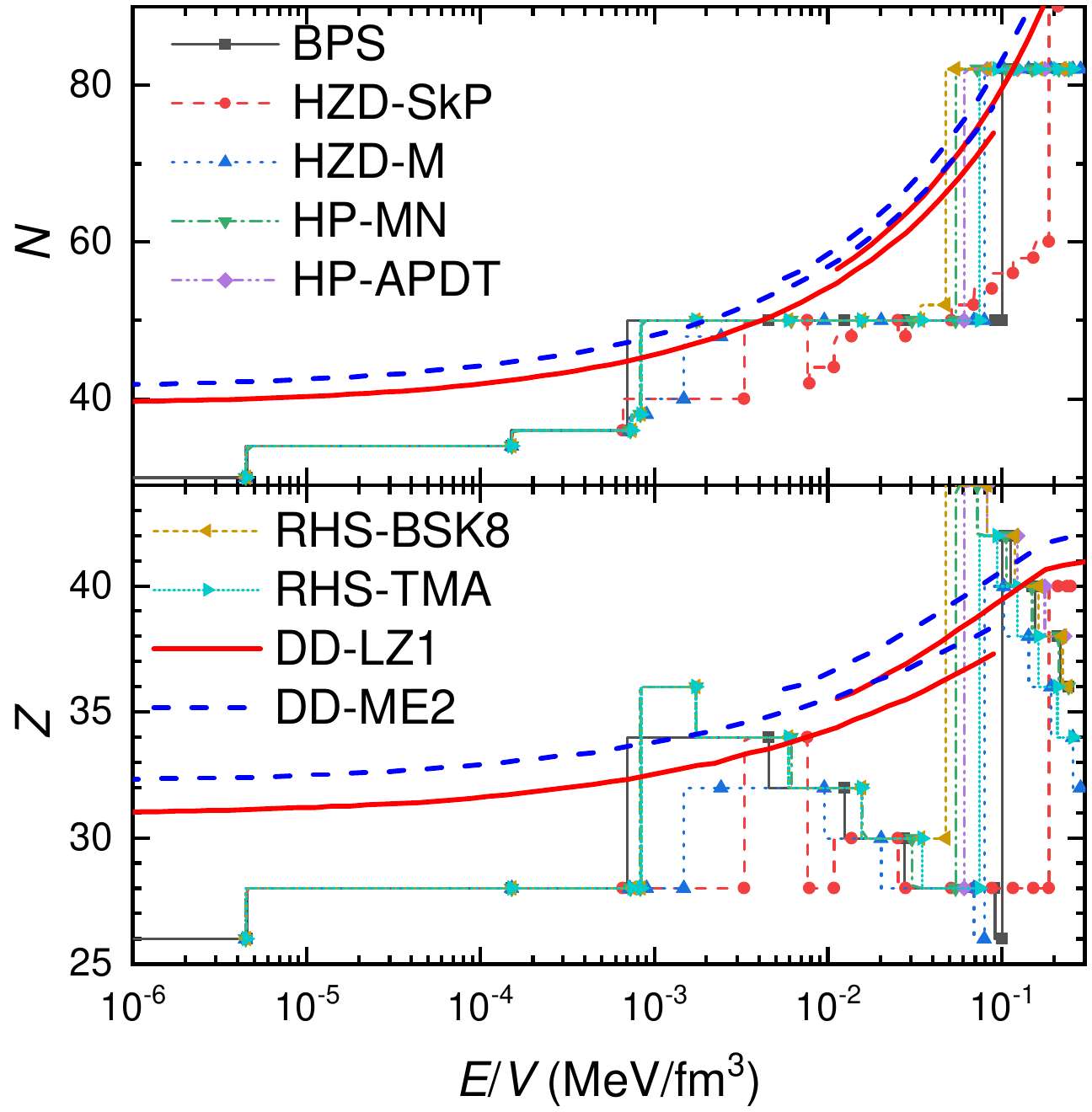}
  \caption{\label{Fig:Nucloen}Neutron ($N$) and proton ($Z$) numbers of nuclei in the outer crusts of neutron stars, where the results obtained in this work are compared with those of BPS models adopting slightly different binding energies of nuclei predicted by various nuclear models~\cite{Ruester2006_PRC73-035804}.}
\end{figure}

At $n_\mathrm{b}\lesssim 0.01$ fm${}^{-3}$, the nuclear interaction has little impact on the microscopic structures of neutron star matter, where the proton number $Z$, WS cell radius $R_\mathrm{W}$, and droplet size $R_\mathrm{d}$ obtained with the two functionals generally coincide with each other. As we decrease the density to infinitesimal, we have $Z\rightarrow \sim$31, $R_\mathrm{d}\rightarrow \sim$5.4 fm, and $R_\mathrm{W}\rightarrow \infty$ for both functionals, which can be attributed to the similar proton fractions in Fig.~\ref{Fig:EOS-beta} and WS cell radii in Fig.~\ref{Fig:Micro-beta}.
To show this explicitly, in Fig.~\ref{Fig:Nucloen} we present the nucleon numbers for nuclei in the outer crusts of neutron stars as functions of energy density $E/V$, where both functionals predict similar numbers that increase with density. The consequence of dividing a WS cell into two parts ($R_\mathrm{in} = 35.84$ fm) for $\beta$-equilibrium matter in the outer crusts can be identified in the overlapped region at $E/V\approx0.01$-0.1 MeV/fm$^3$, where the full calculation predicts slightly larger nucleon numbers. At smaller densities, it is found that the deviations in nucleon numbers caused by dividing a WS cell into two are reduced slightly, where the electron density becomes too small to have any sizable impact on the properties of nuclei. Similar trend is observed for symmetric nuclear matter as well according to Fig.~\ref{Fig:Charge_aprx}. We further compare the sequences of nuclei with those predicted by BPS model, where the binding energies of nuclei obtained with various nuclear models were adopted, i.e., those calculated by
\begin{enumerate}
  \item Baym, Pethick, and Sutherland (BPS)~\cite{Baym1971_ApJ170-299} using the nuclear data of Myers and Swiatecki~\cite{Myers1966_NP81-1};
  \item Haensel, Zdunik, and Dobaczewski~\cite{Haensel1989_AA222-353} using the nuclear data of Dobaczewski, Flocard, and Treiner (HZD-SkP)~\cite{Dobaczewski1984_NPA422-103} and Myers (HZD-M)~\cite{Myers1977};
  \item Haensel and Pichon~\cite{Haensel1994_AA283-313} using the nuclear data of M\"oller and Nix (HP-MN)~\cite{Moller1988_ADNDT39-213} and Aboussir et al. (HP-APDT)~\cite{Aboussir1992_NPA549-155};
  \item R\"uster, Hempel, and Schaffner-Bielich~\cite{Ruester2006_PRC73-035804} using the nuclear data of Skyrme (RHS-BSk8)~\cite{Samyn2004_PRC70-044309} and RMF models (RHS-TMA)~\cite{Geng2005_PTP113-785}.
\end{enumerate}
Due to the lack of experimental data, there are discrepancies on the binding energies for nuclei in outer crusts at $E/V\gtrsim 0.001$ MeV/fm$^3$, which lead to different sequences of nuclei. Since TFA is adopted in our calculation, in contrast to BPS model with discrete nucleon numbers, the nucleon numbers vary smoothly. In general, the values of $Z$ and $N$ obtained here are larger than those of BPS model, while all of them are increasing with density.

In contrast to the cases at small densities, different results are obtained with the two functionals if we examine the density regions at $n_\mathrm{b}\gtrsim 0.01$ fm${}^{-3}$, where DD-LZ1 predicts smaller $R_\mathrm{W}$, larger $Z$ and $R_\mathrm{d}$ as the proton fraction is larger than that of DD-ME2. This is attributed to the differences in the symmetry energy at subsaturation densities as indicated in Fig.~\ref{Fig:Binding}, where the functional DD-LZ1 predicts larger values than that of DD-ME2. The variation in the proton number $Z$ as well as other microscopic structures will affect the transport and elastic properties of neutron star matter, which are essential for interpreting various neutron star observations~\cite{Chamel2008_LRR11-10, Caplan2017_RMP89-041002}. For example, assuming point nucleus embedded in a uniform electron background, the effective shear modulus of a BCC crystal can be estimated with~\cite{Ogata1990_PRA42-4867}
\begin{equation}
  \mu_\mathrm{eff} = 0.1194 \frac{e^2 Z^2}{R_\mathrm{W}V}. \label{eq:mueff_BCC}
\end{equation}
The spectrums of the quasi-periodic oscillations (QPOs) observed after giant flares of soft gamma repeaters are thus expected to be affected by the variation in $Z$ and $R_\mathrm{W}$~\cite{Hansen1980_ApJ238-740, Schumaker1983_MNRAS203-457, McDermott1988_ApJ325-725,  Strohmayer1991_ApJ375-679, Passamonti2012_MNRAS419-638, Gabler2018_MNRAS476-4199, Sotani2012_PRL108-201101, Sotani2016_MNRAS464-3101, Kozhberov2020_MNRAS498-5149}.

\subsection{\label{sec:star} Neutron stars}
\begin{figure}
\includegraphics[width=\linewidth]{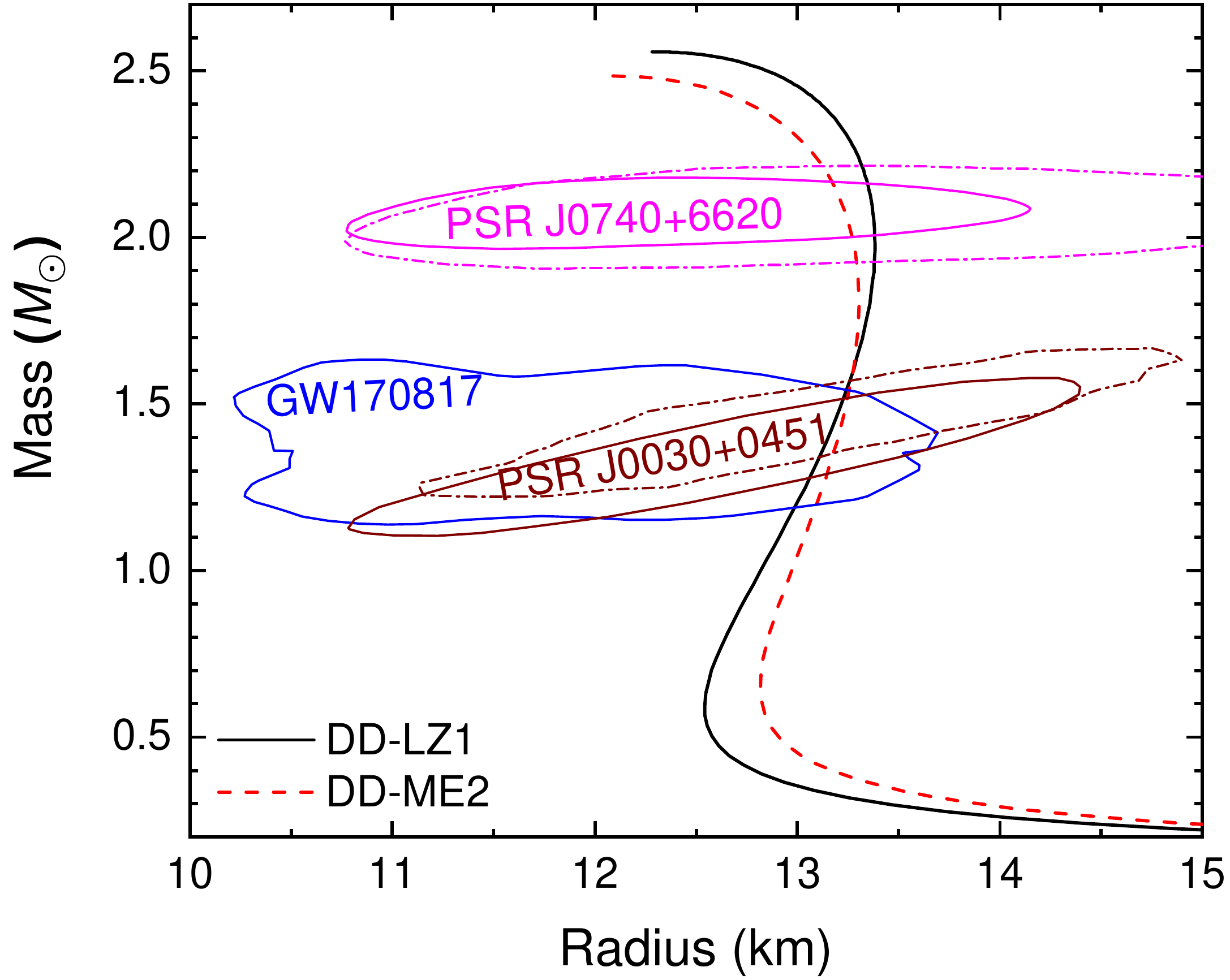}
\caption{\label{Fig:MRL} Mass-radius relations of neutron stars obtained with the two covariant density functionals DD-LZ1~\cite{Wei2020_CPC44-074107} and DD-ME2~\cite{Lalazissis2005_PRC71-024312}. The contours indicate the constraints from the binary neutron star merger event GRB 170817A-GW170817-AT 2017gfo within 90\% credible region~\cite{LVC2018_PRL121-161101}, as well as PSR J0030+0451 and PSR J0740+6620 from NICER and XMM-Newton Data within 68\% credible region ~\cite{Riley2019_ApJ887-L21, Riley2021_ApJ918-L27, Miller2019_ApJ887-L24, Miller2021_ApJ918-L28}.}
\end{figure}

Based on the unified EOSs of neutron star matter presented in Fig.~\ref{Fig:EOS-beta}, we investigate the structures of neutron stars by solving the TOV equation
\begin{eqnarray}
&&\frac{\mbox{d}P}{\mbox{d}r} = -\frac{G M E}{r^2}   \frac{(1+P/E)(1+4\pi r^3 P/M)} {1-2G M/r},  \label{eq:TOV}\\
&&\frac{\mbox{d}M}{\mbox{d}r} = 4\pi E r^2, \label{eq:m_star}
\end{eqnarray}
where the gravity constant $G=6.707\times 10^{-45}\ \mathrm{MeV}^{-2}$. The mass-radius relations of neutron stars predicted by the two functionals are then presented in Fig.~\ref{Fig:MRL}. The corresponding constraints from pulsar observations are indicated as well, i.e., the constraints from the binary neutron star merger event GRB 170817A-GW170817-AT 2017gfo within 90\% credible region~\cite{LVC2018_PRL121-161101}, the constraints of PSR J0030+0451 and PSR J0740+6620 from NICER and XMM-Newton Data plotted in solid~\cite{Riley2019_ApJ887-L21, Riley2021_ApJ918-L27} and dashed~\cite{Miller2019_ApJ887-L24, Miller2021_ApJ918-L28} contours covering 68\% credible region.

It is evident that both the maximum masses obtained with the two functionals easily surpass the two-solar-mass limit of PSR J0740+6620~\cite{Fonseca2021_ApJ915-L12}, where $M_\mathrm{max} = 2.56$ and 2.48 $M_{\odot}$ for DD-LZ1 and DD-ME2, respectively. The radii obtained with the two functionals are slightly different, which is attributed the variations in the stiffness of the EOSs in Fig.~\ref{Fig:EOS-beta}. For a fixed neutron star mass with the center density $n_\mathrm{b}\lesssim 0.3$ fm${}^{-3}$, the radius predicted by DD-LZ1 is smaller than that of DD-ME2 due to the softer EOS in Fig.~\ref{Fig:EOS-beta}. For neutron stars with larger masses, the situation reverses since DD-LZ1 predicts stiffer EOS. In such cases, the combination of small radii/masses and large radii/masses for DD-LZ1 suggests that the neutron star EOS is soft at small densities and stiff at larger densities, which is partly due to the larger curvature parameter $K_\mathrm{sym}$ of symmetric energy and is attributed to the peculiar density dependent behavior of the coupling strengths as discussed in Sec.~\ref{sec:pasta_beta}. More accurate measurements on the radii are necessary in order to tell the difference between the predictions of the two functionals, which could in principle measure $K_\mathrm{sym}$ as well~\cite{Li2021_Universe7-182, Zhang2019_EPJA55-39}.

Finally, it is worth mentioning that the densities at the center of the most massive neutron stars reach $\sim$0.8 fm${}^{-3}$. At such large densities, new degrees of freedom such as mesons ($\pi$, $K$, etc.), heavy baryons ($\Delta$, $\Lambda$, $\Sigma$, $\Xi$, $\Omega$, etc.), and deconfinement phase transition into quarks ($u$, $d$, $s$) may take place, which would effectively reduce the energy density of stellar matter. Consequently, the EOSs of stellar matter becomes softer and the corresponding radii of compact stars become smaller~\cite{Baym2018_RPP81-056902, Sun2019_PRD99-023004, Xia2020_PRD102-023031, Dexheimer2021_PRC103-025808, Sun2021_PRD103-103003}. The possible existence of hyperons adopting the covariant density functionals DD-LZ1 and DD-ME2 are investigated in Refs.~\cite{Sun2022, Tu2022_ApJ925-16}, while other possible scenarios will be examined in our future study.

\section{\label{sec:con}Conclusion}
In this work we have developed a new numerical recipe to investigate the properties of nuclear matter in a unified manner, which covers a wide range of densities with $10^{-10}$ fm${}^{-3} \lesssim n_\mathrm{b} \lesssim 2$ fm${}^{-3}$. The Thomas-Fermi approximation was adopted, where spherical and cylindrical symmetries were assumed for the WS cells. The effects of charge screening was shown to affect the microscopic structures (shape, nuclear radius $R_d$, cell size $R_\mathrm{W}$, etc.) of nuclear pasta~\cite{Maruyama2005_PRC72-015802}. In such cases, we have included the effects of charge screening around the nucleus, where electrons move freely with the density profiles dominated by the Coulomb potential. For fixed nuclear shape, baryon number density $n_\mathrm{b}$, and proton fraction $Y_p$, the optimum WS cell size $R_\mathrm{W}$ was obtained by minimizing the energy of the system, while the ground state was fixed by searching for the minimum energy per baryon among various nuclear shapes.

We then investigate the EOSs of nuclear matter as well as the corresponding microscopic structures adopting a novel relativistic mean field Lagrangian (DD-LZ1) with peculiar density-dependent meson-nucleon couplings, which were compared with that of DD-ME2.
The couplings $g_\sigma$ and $g_\omega$ as functions of density are in parallel to each other for DD-ME2, while this is not the case for DD-LZ1 in order to restore the pseudospin symmetry of the high-$l$ orbitals in finite nuclei. Both functionals predict similar saturation properties for nuclear matter, so that the differences at densities and isospin asymmetries away from $n_\mathrm{b}=n_0$ and $Y_p=0.5$ are mainly due to the different density-dependent behavior of meson-nucleon couplings, which were investigated for the properties of nuclear matter and neutron stars. Various scenarios with both fixed proton fractions and $\beta$-equilibration were examined. It was found that typical nuclear matter structures (droplet, rod, slab, tube, bubble, and uniform) emerge sequentially as density increases. The results obtained with the two covariant density functionals generally coincide with each other, while the density range for nonuniform nuclear matter obtained with DD-ME2 is slightly larger than that of DD-LZ1. For neutron star matter in $\beta$-equilibrium, the two functionals also predict similar results throughout the density range. Nevertheless, the differences on the microscopic structures of neutron star matter are evident at $n_\mathrm{b}\gtrsim 0.01$ fm${}^{-3}$, where DD-LZ1 predicts smaller $R_\mathrm{W}$, larger $Z$ and $R_\mathrm{d}$ as the proton fraction is larger than that of DD-ME2. The variation in $Z$, $R_\mathrm{d}$, and $R_\mathrm{W}$ would affect the transport and elastic properties of neutron star matter, which are expected to alter various physical processes in neutron star properties and evolutions~\cite{Chamel2008_LRR11-10, Caplan2017_RMP89-041002}, e.g., the spectrums of the QPOs~\cite{Hansen1980_ApJ238-740, Schumaker1983_MNRAS203-457, McDermott1988_ApJ325-725,  Strohmayer1991_ApJ375-679, Passamonti2012_MNRAS419-638, Gabler2018_MNRAS476-4199, Sotani2012_PRL108-201101, Sotani2016_MNRAS464-3101, Kozhberov2020_MNRAS498-5149}, the release of magnetic and elastic energy observed in magnetar bursts~\cite{Beloborodov2014_ApJ794-L24, Beloborodov2016_ApJ833-261, Li2016_ApJ833-189}, the short gamma-ray burst precursors of neutron star mergers~\cite{Tsang2012_PRL108-011102}, pulsar glitches~\cite{Ruderm1969_Nature223-597, Baym1971_AP66-816, Haskell2015_IJMPD24-1530008, Akbal2017_MNRAS473-621, Guegercinoglu2019_MNRAS488-2275, Layek2020_MNRAS499-455, Shang2021_ApJ923-108}, and the gravitational waves emitted by fast rotating neutron stars~\cite{Abbott2020_ApJ902-L21}. Meanwhile, the EOS obtained with DD-LZ1 is softer than that of DD-ME2 at $n_\mathrm{b}\lesssim 0.3$ fm${}^{-3}$, which becomes stiffer at larger densities. This is due to the larger curvature parameter $K_\mathrm{sym}$ of symmetry energy for DD-LZ1, which is attributed to the peculiar density dependent behavior of the coupling strengths. The variations of the EOSs have a direct consequence on the mass-radius relations of neutron stars, where the radii of neutron stars predicted by DD-LZ1 are smaller than that of DD-ME2 at masses lower than $1.6 M_{\odot}$ but larger for more massive neutron stars.

\section*{ACKNOWLEDGMENTS}
We would like to thank Professor Nobutoshi Yasutake and Professor Toshitaka Tatsumi for fruitful discussions. This work was supported by the National SKA Program of China (Grant No.~2020SKA0120300), the National Natural Science Foundation of China (Grants No.~11875052, No.~11873040, No.~11705163, and No.~11525524), the Fundamental Research Funds for the Central Universities (Grant No.~lzujbky-2021-sp36), the science research grants from the China Manned Space Project (Grant No.~CMS-CSST-2021-B11), the Youth Innovation Fund of Xiamen (Grant  No.~3502Z20206061), and the National Key R\&D Program of China (Grant No.~2018YFA0404402).


\newpage

%

\end{document}